\documentclass[twocolumn,english,aps,reprint]{revtex4}
\usepackage[T1]{fontenc}
\usepackage[latin9]{inputenc}
\usepackage{amsmath}
\usepackage{graphicx}
\usepackage{amssymb}
\usepackage{esint}

\makeatletter

\providecommand{\tabularnewline}{\\}
\newcommand{\lyxdot}{.}

\@ifundefined{textcolor}{}
{%
 \definecolor{BLACK}{gray}{0}
 \definecolor{WHITE}{gray}{1}
 \definecolor{RED}{rgb}{1,0,0}
 \definecolor{GREEN}{rgb}{0,1,0}
 \definecolor{BLUE}{rgb}{0,0,1}
 \definecolor{CYAN}{cmyk}{1,0,0,0}
 \definecolor{MAGENTA}{cmyk}{0,1,0,0}
 \definecolor{YELLOW}{cmyk}{0,0,1,0}
 }

\makeatother

\makeatother

\usepackage{babel}

\makeatother

\usepackage{babel}

\makeatother

\usepackage{babel}

\begin{document}

\title{$U(1)$ slave-particle study of the finite-temperature doped Hubbard
model in one and two dimensions }

\author{P. Ribeiro}

\email{ribeiro@cfif.ist.utl.pt}

\affiliation{CFIF, Instituto Superior Técnico, Universidade Técnica de Lisboa,
\\
 Av. Rovisco Pais, 1049-001 Lisboa, Portugal}

\author{P. D. Sacramento}

\email{pdss@cfif.ist.utl.pt}

\affiliation{CFIF, Instituto Superior Técnico, Universidade Técnica de Lisboa,
\\
 Av. Rovisco Pais, 1049-001 Lisboa, Portugal}

\author{M. A. N. Araújo}

\email{mana@evunix.uevora.pt}

\affiliation{CFIF, Instituto Superior Técnico, Universidade Técnica de Lisboa,
\\
 Av. Rovisco Pais, 1049-001 Lisboa, Portugal}

\affiliation{Departamento de Física, Universidade de Évora, P-7000-671 Évora,
Portugal}
\begin{abstract}
One-dimensional systems have unusual properties such as fractionalization
of degrees of freedom. Possible extensions to higher dimensional systems
have been considered in the literature. In this work we construct
a mean field theory of the Hubbard model taking into account a separation
of the degrees of freedom inspired by the one-dimensional case and
study the finite-temperature phase diagram for the Hubbard chain and
square lattice. The mean field variables are defined along the links
of the underlying lattice. We obtain the spectral function and identify
the regions of higher spectral weight with the fractionalized fermionic
(spin) and bosonic (charge) excitations. 
\end{abstract}
\maketitle
\global\long\def\ket#1{\left| #1\right\rangle }

\global\long\def\bra#1{\left\langle #1 \right|}

\global\long\def\kket#1{\left\Vert #1\right\rangle }

\global\long\def\bbra#1{\left\langle #1\right\Vert }

\global\long\def\braket#1#2{\left\langle #1\right. \left| #2 \right\rangle }

\global\long\def\bbrakket#1#2{\left\langle #1\right. \left\Vert #2\right\rangle }

\global\long\def\av#1{\left\langle #1 \right\rangle }

\global\long\def\tr{\text{Tr}}

\global\long\def\im{\text{Im}}

\global\long\def\re{\text{Re}}

\global\long\def\sign{\text{sign}}

\section{introduction}

The Hubbard model is one of the simplest models accounting for interactions
between electrons and has been used to describe various strongly correlated
materials. It is parameterized by only two constants, a kinetic term
scaled in the tight-binding approximation by the hopping amplitude
$t$ and an on-site interaction $U$ modeling the Coulomb repulsion
felt by two electrons of opposite spins occupying the same site.

The exact solution of the one-dimensional Hubbard model by the Bethe
ansatz \cite{Lieb_1968} involves composite degrees of freedom that
correspond to different rapidity branches. In addition to a $c0$
charge-momentum rapidity, there are sets of $(\alpha,\nu)$ rapidities
\cite{Takahashi_1972}. The general rapidity branch label $\alpha\nu$
is such that $\alpha=c,s$ and $\nu=0,1,2,...$ for $\alpha=c$ and
$\nu=1,2,...$ for $\alpha=s$. The $(c,\nu)$ and $(s,\nu)$ rapidities
are associated with the charge and spin degrees of freedom, respectively.
For electronic densities $n\leq1$, the ground state has finite occupancies
for the charge $(c,0)$ and spin $(s,1)$ branches only. The relation
between the original electrons and the entities that describe the
eigenstates of the Hubbard model is however complex and only recently
some light has emerged \cite{Carmelo_2004}.

Electron double occupancy is a good quantum number (it is conserved)
in the limit of $U\rightarrow\infty$ but for finite values of $U/t$
it is not conserved. However, it is possible to define new fermionic
operators, associated with fermionic objects called rotated electrons,
through a canonical transformation, $\hat{V}$, such that the double
occupancy of these rotated electrons is a good quantum number for
all finite values of $U/t$ \cite{Harris_1967,Carmelo_2004,Stein_1997,MacDonald_1988}.
In terms of the rotated electrons, it is a consistent interpretation
of the Bethe ansatz states to describe the various branches (rapidities)
in terms of, first, a separation into empty, double-occupied and singly
occupied sites. The empty and double-occupied sites are called $\eta$-spin
$1/2$ holons \cite{Carmelo_2004} and the singly-occupied give rise
to a charge part that originates the $(c,0)$ particles and to a spin
part which originates the spin-$1/2$ spinons. The $\eta$-spin projection
$1/2$ (and $-1/2$) holons correspond to rotated-electron unoccupied
sites (and doubly-occupied sites). The spinons of spin projection
$\pm1/2$ refer to the spins of the rotated electrons which singly
occupied sites. Second, these are paired in singlets ($s,1$) or in
pairs of pairs of singlets and so on ($s,\nu$). The holons are also
paired in such a way that empty sites and doubly-occupied sites are
paired ($c,\nu$). These results apply to the rotated electrons and
not to the original electrons. They are however related by the above
mentioned canonical transformation. At high values of $U$ they are
very close and identical when $U\rightarrow\infty$. For many practical
situations, such as the calculation of various correlation functions,
it is a reasonably good approximation to consider the rotated electrons
as similar to the original electrons \cite{Carmelo_2000,Sing_2003,Carmelo_2004_b,Carmelo_2006,Bozi_2008}.

There are transformations in the literature that propose a similar
decoupling of the electronic degrees of freedom. This can be seen
for instance in \cite{Kotliar_1986} or in \cite{Zou_1988}. The main
motivation was the study of either the large-$U$ limit in the Hubbard
or Anderson models \cite{Dorin_1992} with the intent to control in
an efficient way the projection to states where double occupancy is
restricted (as in the $t-J$ model) but considering a finite value
of $U$ instead of the extreme case of infinite $U$, usually taken
care of by a single slave boson \cite{Coleman_1984}. In the Zou-Anderson
transformation (ZA) each physical electron is mapped into the one
particle sector of a set of four particles, two fermions and two bosons.
The two bosons may be chosen as spinless particles and represent the
empty and doubly occupied states of each lattice site; excitations
of these two degrees of freedom are called holons and doublons and
can be interpreted as carrying no charge or $-2e$ respectively. The
fermions then represent singly occupied states with spin $\pm1/2$,
excitations on this sector are called spinons and carry charge $-e$.
However, the charges may be defined differently, as considered in
the original paper by Zou and Anderson \cite{Zou_1988}. There is
also an exact transformation that explicitly includes spin and charge
separation \cite{Ostlund_2006} introducing two sets of operators
as quasicharge (fermionic) and quasispin (spin-like). The representation
of the quasispin operators leads however in general to bilinear terms
in bosons or fermions and therefore this leads to six operator terms
in the Hamiltonian.

The ZA mapping reverses the role of the interacting and kinetic terms
in the Hamiltonian. The interacting Hubbard term becomes quadratic
in the ZA particles and the kinetic one is transformed into an interacting
quartic term that couples particles along the lattice links.  This
is particularly useful to study the strong interacting (large U) regime
where the kinetic term is treated as a perturbation. The price of
this transformation is the appearance of an on-site constraint which
assures exactly one particle per lattice site. In the mean field (MF)
approach this translates to an on-site Lagrange multiplier. Generically
slave-particle methods induce new unphysical symmetries of the Hamiltonian
written in terms of slave particles, that are called in this context
gauge symmetries. In this particular case the symmetry group is U(1).

We note that the representation introduced by Zou and Anderson has
been used to explicitly obtain an exact solution of the Hubbard chain
in the large $U$ limit in a much simpler way as compared to the Bethe
ansatz \cite{Dias_1992}. Also it has been used to study the stiffness
of the one-dimensional Hubbard model in a way equivalent and complementar
to the Bethe ansatz solution \cite{Peres_2000}.

In the large U limit it is very costly to create doubly occupied sites.
It is usual to consider $d=0$ \cite{Lee_2006} (projection to subspace
of no-double occupancy) or to perform a canonical transformation to
eliminate transitions to doubly occupied sites \cite{Zou_1988}. This
type of procedure leads to a treatment similar to the $t-J$ model,
appropriate near half filling in the vicinity of magnetic order for
a square lattice.

It has been argued recently that one should instead integrate out
the high-energy scale to obtain an effective low-energy theory which
has been shown to contain a charge $2e$ bosonic mode \cite{Leigh_2007,Choy_2008,Phillips_2009}.
This mode may be bound to a hole, providing a possible explanation
of the pseudogap observed in high-$T_{c}$ materials. In this work
we will maintain the full structure of the transformation between
the electron operators and the auxiliary particles introduced in the
Zou-Anderson transformation. We will be focusing on the finite energy
(finite temperature) properties where we will find and study phases
with non-standard correlation functions associated with the fractionalized
degrees of freedom. We will obtain the phase diagrams for one- and
two-dimensional systems and calculate various correlation functions
as well as the spectral function.

We will use nonlocal decouplings of the auxiliary operators leading
to link variables that can be associated with nearest-neighbor spin
singlets or bound-states between empty and doubly occupied sites in
a way close to the Bethe ansatz solution and also suggested by the
treatment of Ref. \cite{Phillips_2009}, imposing at MF level the
existence of these states. A simplified treatment of link variables
was introduced in Ref. \cite{Alvarez_1995} and the significance of
short-range correlations between empty and doubly occupied sites was,
for instance, determined in Ref. \cite{Kaplan_1982}. We note that
bond variables appear naturally in extensions of the Hubbard model
too, for instance nearest-neighbor interactions, or bond correlated
hoppings \cite{Japaridze_1999}. Interestingly, many results have
been obtained for systems where the hopping between singly
occupied sites and empty and doubly occupied sites is eliminated,
implying that double occupancy is a good quantum number \cite{Strack_1993,Arrachea_1994,Dolcini_2002}.

The paper is organized as follows: in section II we discuss the methods
used to construct the MF solution. In section III we discuss the MF
phase diagram for the Hubbard chain and the square lattice and in
section IV we discuss the spectral function for both cases, interpreted
in terms of the fractional excitations here considered. Finally in
section V we discuss the results obtained and in the appendix we present
a list of MF solutions for the square lattice.

\section{methods}

We start from the $U(1)$ representation of the Hubbard model introduced
by Zou and Anderson (ZA) \cite{Zou_1988}, $c_{r,\sigma}^{\dagger}=e_{r}\, s_{r,\sigma}^{\dagger}+\sigma d_{r}^{\dagger}\, s_{r,-\sigma}$
which maps a spin-$1/2$ fermion (the physical electron) to the one-particle
subspace of four fields $e_{r},d_{r},s_{r,\sigma=\pm1}$. In this
work bosonic $e_{r}$ and $d_{r}$ fields (where $r$ labels the lattice
sites) are considered corresponding to the annihilation of empty and
doubly occupied sites, $s_{r,\sigma=\pm1}$ are fermions that carry
the spin degree of freedom. The opposite choice of statistics ($e,d$
fermions and $s_{\pm1}$ bosons) is also possible leaving the mapping
unchanged. The Hamiltonian of the Hubbard model is given by \begin{widetext}
\begin{eqnarray}
H-\mu\, n_{T} & = & -t\sum_{r,\delta>0,\sigma}c_{r,\sigma}^{\dagger}c_{r+\delta,\sigma}+c_{r+\delta,\sigma}^{\dagger}c_{r,\sigma}+U\sum_{r}n_{r,1}n_{r,-1}-\mu\sum_{r,\sigma}n_{r,\sigma},\label{eq:Ham}\\
 & = & t\sum_{r,\delta>0}\left[\left(d_{r}^{\dagger}d_{r+\delta}-e_{r}^{\dagger}e_{r+\delta}\right)^{\dagger}\left(s_{r,1}^{\dagger}s_{r+\delta,1}+s_{r-1}^{\dagger}s_{r+\delta,-1}\right)+\left(s_{r,1}^{\dagger}s_{r+\delta,1}+s_{r-1}^{\dagger}s_{r+\delta,-1}\right)^{\dagger}\left(d_{r}^{\dagger}d_{r+\delta}-e_{r}^{\dagger}e_{r+\delta}\right)\right.\nonumber \\
 &  & \left.+\left(d_{r}e_{r+\delta}+e_{r}d_{r+\delta}\right)^{\dagger}\left(s_{r,1}s_{r+\delta,-1}-s_{r,-1}s_{r+\delta,1}\right)+\left(s_{r,1}s_{r+\delta,-1}-s_{r,-1}s_{r+\delta,1}\right)^{\dagger}\left(d_{r}e_{r+\delta}+e_{r}d_{r+\delta}\right)\right]\nonumber \\
 &  & +U\sum_{r}d_{r}^{\dagger}d_{r}-\mu\sum_{r}(1+d_{r}^{\dagger}d_{r}-e_{r}^{\dagger}e_{r})\end{eqnarray}
 \end{widetext} where $\delta$ is a directed lattice vector connecting
nearest neighbor sites, $\mu$ is the chemical potential and $n_{T}$
the total number of electrons.\\
 \\
 Using the ZA mapping the partition function is given in a path
integral formulation by \begin{eqnarray}
Z & = & \int D\lambda DeDdDs_{\sigma}\ e^{-\int_{0}^{\beta}\ d\tau\, L}\\
L & = & \sum_{r}d_{r}^{\dagger}(\partial_{\tau}+\lambda_{r}+U-\mu)d_{r}+\nonumber \\
 &  & \sum_{r}e_{r}^{\dagger}(\partial_{\tau}+\lambda_{r}+\mu)e_{r}+\sum_{r}s_{r,\sigma}^{\dagger}(\partial_{\tau}+\lambda_{r})s_{r,\sigma}\nonumber \\
 &  & +\sum_{r,\delta}t_{\delta}\, tr\left[\mathbf{B}_{r,r+\delta}.\mathbf{F}_{r,r+\delta}\right]-\sum_{r}\lambda_{r}-\mu V\end{eqnarray}
 where \begin{eqnarray}
\mathbf{B}_{i,j} & = & \left(\begin{array}{c|c}
\mathit{d}_{j}^{\dagger}\text{ }\mathit{d}_{i}-\mathit{e}_{j}^{\dagger}\mathit{e}_{i} & \mathit{d}_{i}^{\dagger}\mathit{e}_{j}^{\dagger}+\mathit{d}_{j}^{\dagger}\mathit{e}_{i}^{\dagger}\\
\hline \mathit{e}_{i}\mathit{d}_{j}+\mathit{e}_{j}\mathit{d}_{i} & \mathit{d}_{i}^{\dagger}\mathit{d}_{j}-\mathit{e}_{i}^{\dagger}\mathit{e}_{j}\end{array}\right);\nonumber \\
\mathbf{F}_{i,j} & = & \left(\begin{array}{c|c}
\mathit{s}_{i,-1}^{\dagger}\mathit{s}_{j,-1}+\mathit{s}_{i,1}^{\dagger}\mathit{s}_{j,1} & \mathit{s}_{i,-1}^{\dagger}\mathit{s}_{j,1}^{\dagger}-\mathit{s}_{i,1}^{\dagger}\mathit{s}_{j,-1}^{\dagger}\\
\hline \mathit{s}_{j,1}\mathit{s}_{i,-1}-\mathit{s}_{j,-1}\mathit{s}_{i,1} & \mathit{s}_{j,-1}^{\dagger}\mathit{s}_{i,-1}+\mathit{s}_{j,1}^{\dagger}\mathit{s}_{i,1}\end{array}\right);\nonumber \\
\end{eqnarray}
 are respectively bosonic and fermionic matrices, $\lambda_{r}$ is
an on-site real field inserted in order to project to the one-particle
sector $n_{r}^{\text{ZA}}=d_{r}^{\dagger}d_{r}+e_{r}^{\dagger}e_{r}+\sum_{\sigma}s_{r,\sigma}^{\dagger}s_{r,\sigma}=1$
with$\int D\lambda=\int_{0-i\,\pi/\beta}^{0+i\,\pi/\beta}\prod_{r}\prod_{\tau}i\frac{d\lambda_{r}(\tau)}{2\pi\beta^{-1}}$,
and $V$ is the total number of lattice sites. Using the identity:
$\int d\mu\left(\boldsymbol{Q}^{\dagger},\boldsymbol{Q}\right)e^{-\tr\left[\boldsymbol{Q}^{\dagger}.\boldsymbol{Q}+\boldsymbol{C}^{\dagger}.\boldsymbol{Q}+\boldsymbol{Q}^{\dagger}.\boldsymbol{A}\right]}=e^{\tr\left[\boldsymbol{C}^{\dagger}.\boldsymbol{A}\right]},$
with $d\mu\left(\boldsymbol{Q}^{\dagger},\boldsymbol{Q}\right)=\prod_{i,j}\frac{d\bar{Q}_{j,i}dQ_{i,j}}{2\pi i}$
and $\left[\boldsymbol{Q}^{\dagger}\right]_{i,j}=\bar{Q}_{i,j}$,
$\left[\boldsymbol{Q}\right]_{i,j}=Q_{i,j}$($\bar{Q}_{j,i}=Q_{i,j}^{*}$),
we introduce a $2\times2$ matricial Hubbard-Stratonovich (HS) field
in order to decouple the fermionic and bosonic terms: \begin{eqnarray}
Z & = & \int D\lambda DeDdDs_{\sigma}D\boldsymbol{Q}\ e^{-\int_{0}^{\beta}\ d\tau\, L}\\
\nonumber \\L & = & \sum_{r}d_{r}^{\dagger}(\partial_{\tau}+\lambda_{r}+U-\mu)d_{r}+\sum_{r}e_{r}^{\dagger}(\partial_{\tau}+\lambda_{r}+\mu)e_{r}\nonumber \\
 &  & +\sum_{r}s_{r,\sigma}^{\dagger}(\partial_{\tau}+\lambda_{r})s_{r,\sigma}-\sum_{r}\lambda_{r}\nonumber \\
 &  & -\mu V+\sum_{r,\delta}\tr\left[\boldsymbol{Q}_{r,\delta}^{\dagger}.\boldsymbol{Q}_{r,\delta}\right]\nonumber \\
 &  & +\sum_{r,\delta}\sqrt{t_{\delta}}\tr\left[\boldsymbol{Q}_{r,\delta}^{\dagger}.\mathbf{F}_{r,r+\delta}-\mathbf{B}_{r,r+\delta}.\boldsymbol{Q}_{r,\delta}\right].\label{eq:H-S-Lag}\end{eqnarray}
 The additional gauge freedom introduced when writing the Hubbard
model in this particular slave-particle form is $U(1)$ and is implemented
by the operator $U(\phi)=\prod_{r}e^{-i\phi_{r}n_{T,r}}$. Such gauge
transformation changes the ZA particle fields by a site dependent
phase $a_{r}\to e^{-i\phi_{r}}a_{r}$ ($a=s_{\pm},d,e$) which gives
the simple transformation rule for the matrices $\mathbf{B}_{r,r+\delta}\to e^{-i\sigma_{z}\phi_{r+\delta}}.\mathbf{B}_{r,r+\delta}.e^{i\sigma_{z}\phi_{r}}$
and $\mathbf{F}_{r,r+\delta}\to e^{-i\sigma_{z}\phi_{r}}.\mathbf{F}_{r,r+\delta}.e^{i\sigma_{z}\phi_{r+\delta}}$.
In order to leave the Lagrangian invariant, this transformation also
induces a gauge transformation in the HS fields \begin{eqnarray}
\boldsymbol{Q}_{r,\delta} & \to & e^{-i\sigma_{z}\phi_{r}}.\mathbf{Q}_{r,\delta}.e^{i\sigma_{z}\phi_{r+\delta}},\end{eqnarray}
 defining also 6 on-site gauge invariant quantities: $\left[\boldsymbol{Q}_{r,\delta}\right]_{i,j}\left[\boldsymbol{Q}_{r,\delta}^{\dagger}\right]_{i,j}$
and $\left[\boldsymbol{Q}_{r,\delta}\right]_{i,j}\left[\boldsymbol{Q}_{r,\delta}\right]_{j,i}$($i,j$=1,2).

A MF treatment of this Lagrangian can be justified introducing $N$
copies of the ZA particles in order to perform a $1/N$ expansion
that coincides with the usual MF approximation at zero order and organizes
the following corrections in numbers of loops\cite{Lee_2006}. However,
as in $SU(N)$ generalizations of spin-1/2 models, this parameter
is rather unphysical and this approach will not be explicitly pursued
here. The MF approximation is obtained varying the free energy with
respect to the $\boldsymbol{Q}$ and $\lambda$ fields: \begin{eqnarray}
\boldsymbol{Q}_{r,\delta}(\tau) & = & -\sqrt{t_{\delta}}\av{\mathbf{F}_{r,\delta}(\tau)}_{0}\label{eq:MF-1}\\
\boldsymbol{Q}_{r,\delta}^{\dagger}(\tau) & = & \sqrt{t_{\delta}}\av{\mathbf{B}_{r,\delta}(\tau)}_{0}\label{eq:MF-2}\\
1 & = & \av{n^{ZA}(\tau))}_{0},\label{eq:MF-3}\end{eqnarray}
 where $\av{\ }_{0}$ stands for the MF average. In order to study
this set of equations a time and space translational invariant ansatz
$\boldsymbol{Q}_{r,\delta}(\tau)=\boldsymbol{Q}_{\delta}$ was imposed.
The saddle-point values of the HS fields describe hopping and pairing
terms:\begin{eqnarray}
\chi_{F,r,\delta} & = & \av{\mathit{s}_{r+\delta,1}^{\dagger}\mathit{s}_{r,1}+\mathit{s}_{r+\delta,-1}^{\dagger}s_{r,-1}}_{0},\nonumber \\
\Delta_{F,r,\delta} & = & \av{\mathit{s}_{r+\delta,1}\mathit{s}_{r,-1}-\mathit{s}_{r+\delta,-1}\mathit{s}_{r,1}}_{0},\nonumber \\
\chi_{B,r,\delta} & = & \av{\mathit{d}_{r+\delta}^{\dagger}\mathit{d}_{r}-\mathit{e}_{r+\delta}^{\dagger}\mathit{e}_{r}}_{0},\nonumber \\
\Delta_{B,r,\delta} & = & \av{\mathit{d}_{r+\delta}\mathit{e}_{r}+\mathit{e}_{r+\delta}\mathit{d}_{r}}_{0},\label{eq:chi_Delta}\end{eqnarray}
 and Eqs. (\ref{eq:MF-1}-\ref{eq:MF-2}) are equivalent to \begin{eqnarray}
\boldsymbol{Q}_{r,\delta} & = & -\sqrt{t_{\delta}}\left(\begin{array}{cc}
\chi_{F,r,\delta}^{\dagger} & \Delta_{F,r,\delta}^{\dagger}\\
\Delta_{F,r,\delta} & \chi_{F,r,\delta}\end{array}\right),\nonumber \\
\boldsymbol{Q}_{r,\delta}^{\dagger} & = & \sqrt{t_{\delta}}\left(\begin{array}{cc}
\chi_{B,r,\delta} & \Delta_{B,i,\delta}^{\dagger}\\
\Delta_{B,r,\delta} & \chi_{B,r,\delta}^{\dagger}\end{array}\right).\end{eqnarray}
 Note that the MF values obtained for $\boldsymbol{Q}^{\dagger}$
and $\boldsymbol{Q}$ are not complex conjugated of each other; this
is crucial in order to interpret the zero order Lagrangian as coming
from a hermitian MF Hamiltonian, in the extended Hilbert space, as
noticed in \cite{Lee_2005,Alvarez_1995}. This corresponds to the
analytic continuation of the $\boldsymbol{Q}$ fields and also occurs
for $\lambda$. Care should be taken, however, when considering fluctuations
of these fields around their MF values: as the $\lambda$ fluctuations
are purely imaginary even if the MF value is real, the conjugated
fluctuations of $\delta\chi_{F,r,\delta}$ are $\delta\chi_{B,r,\delta}^{\dagger}$
and not $\delta\chi_{F,r,\delta}^{\dagger}$ as the MF treatment could
suggest. The filling fraction of the electrons is imposed as usual
requiring the chemical potential to satisfy $x=1-\av{n_{T}}=1-\frac{1}{V\beta}\partial_{\mu}\ln Z$,
where $x$ is the hole doping.

Assuming translational invariance of the MF solutions, the MF Hamiltonian
can be brought to a diagonal form in $k$ space. Defining $\tilde{\boldsymbol{b}}_{k}=\left\{ d_{k},e_{-k,1},d_{k,1}^{\dagger},e_{-k,1}^{\dagger}\right\} ^{T}$and
$\tilde{\boldsymbol{s}}_{k}=\left\{ s_{k,1},s_{-k,-1},s_{k,1}^{\dagger},s_{-k,-1}^{\dagger}\right\} ^{T}$the
Bogoliubov transformation diagonalizing the Lagrangian is given by
\begin{eqnarray}
\boldsymbol{a}_{k} & = & R_{b}(k).\tilde{\boldsymbol{b}}_{k}\nonumber \\
\boldsymbol{f}_{k} & = & R_{s}(k).\tilde{\boldsymbol{s}}_{k}\end{eqnarray}
 where $f_{\sigma=\pm}$ and $a_{i=e,d}$ are the fermionic and bosonic
Bogoliubov transformed fields . In these new variables the Hamiltonian
is given by \begin{eqnarray}
H & = & \sum_{k,i=d,e}\, a_{k,i}^{\dagger}\varepsilon_{i,k}a_{k,i}+\Theta_{B,k}\nonumber \\
 &  & +\sum_{k,\sigma}\, f_{k,\sigma}^{\dagger}\varepsilon_{F,k}f_{k,\sigma}+\Theta_{F,k}\nonumber \\
 &  & -V\lambda+V\sum_{\delta}\tr\left[\boldsymbol{Q}_{\delta}^{\dagger}.\boldsymbol{Q}_{\delta}\right]-V\left(\mu+\frac{U}{2}\right)\end{eqnarray}
 where the single particle energies $\varepsilon$ and the energy
shifts $\Theta$ are given by\begin{eqnarray}
\varepsilon_{F,k} & = & \frac{1}{2}\left(\omega_{F,k}-\omega_{F,-k}+\sqrt{4\delta_{F,k}\bar{\delta}_{F,k}+\left(\omega_{F,-k}+\omega_{F,k}\right)^{2}}\right)\nonumber \\
\Theta_{F,k} & = & -\frac{1}{2}\left[\varepsilon_{F,k}+\varepsilon_{F,-k}\right.\nonumber \\
 &  & +\left.\chi_{B}(k)-\chi_{B}^{\dagger}(k)+\chi_{B}(-k)-\chi_{B}^{\dagger}(-k)\right]\nonumber \\
\varepsilon_{d,k} & = & \frac{1}{2}\left(\omega_{d,k}-\omega_{e,-k}+\sqrt{\left(\omega_{e,-k}+\omega_{d,k}\right)^{2}-4\delta_{B,k}\bar{\delta}_{B,k}}\right)\nonumber \\
\varepsilon_{e,k} & = & \frac{1}{2}\left(\omega_{e,k}-\omega_{d,-k}+\sqrt{\left(\omega_{e,k}+\omega_{d,-k}\right)^{2}-4\delta_{B,k}\bar{\delta}_{B,k}}\right)\nonumber \\
\Theta_{B,k} & = & \frac{1}{2}\left[\varepsilon_{d,k}+\varepsilon_{e,-k}\right.\nonumber \\
 &  & +\left.\chi_{F}(k)-\chi_{F}^{\dagger}(k)-\chi_{F}(-k)+\chi_{F}^{\dagger}(-k)\right]\end{eqnarray}
 with \begin{eqnarray}
\omega_{d,k} & = & U-\mu+\lambda+\chi_{F}(k)+\chi_{F}^{\dagger}(k)\nonumber \\
\omega_{e,-k} & = & \mu+\lambda-\chi_{F}(-k)-\chi_{F}^{\dagger}(-k)\nonumber \\
\delta_{B,k} & = & \Delta_{F}(-k)+\Delta_{F}(k)\nonumber \\
\omega_{F,k} & = & \lambda+\chi_{B}(k)+\chi_{B}^{\dagger}(k)\nonumber \\
\delta_{F,k} & = & \Delta_{B}(-k)+\Delta_{B}(k)\end{eqnarray}
 and $A(k)=\sum_{\delta=out}t_{\delta}e^{ik\delta}A_{\delta}$ ($A=\chi,\Delta$).
Explicitly the Bogoliubov rotation matrices are given by

\begin{eqnarray}
R_{s} & = & \left(\begin{array}{cccc}
u_{F,k} & 0 & 0 & v_{F,k}\\
0 & u_{F,k} & -v_{F,k} & 0\\
0 & \bar{v}_{F,k} & u_{F,k} & 0\\
-\bar{v}_{F,k} & 0 & 0 & u_{F,k}\end{array}\right);\nonumber \\
R_{b} & = & \left(\begin{array}{cccc}
u_{B,k} & 0 & 0 & v_{B,k}\\
0 & u_{B,k} & v_{B,k} & 0\\
0 & \bar{v}_{B,k} & u_{B,k} & 0\\
\bar{v}_{B,k} & 0 & 0 & u_{B,k}\end{array}\right)\end{eqnarray}
 with \begin{eqnarray*}
u_{F/B,k} & = & \frac{1}{\sqrt{1+\tau_{F/B}(k)\bar{\tau}_{F/B}(k)}}\\
v_{F/B,k} & = & \frac{\tau_{F/B}}{\sqrt{1+\tau_{F/B}(k)\bar{\tau}_{F/B}(k)}}\\
\tau_{F}(k) & = & \frac{\left(\omega_{F,k}+\omega_{F,-k}\right)-\sqrt{\left(\omega_{F,k}+\omega_{F,-k}\right)^{2}+4\delta_{F,k}\bar{\delta}_{F,k}}}{2\bar{\delta}_{F,k}}\\
\tau_{B}(k) & = & \frac{\left(\omega_{d,k}+\omega_{e,-k}\right)-\sqrt{\left(\omega_{d,k}+\omega_{e,-k}\right)^{2}-4\delta_{B,k}\bar{\delta}_{B,k}}}{2\bar{\delta}_{B,k}}\end{eqnarray*}

\begin{figure*}
\begin{tabular}{ccc}
\includegraphics[width=5cm]{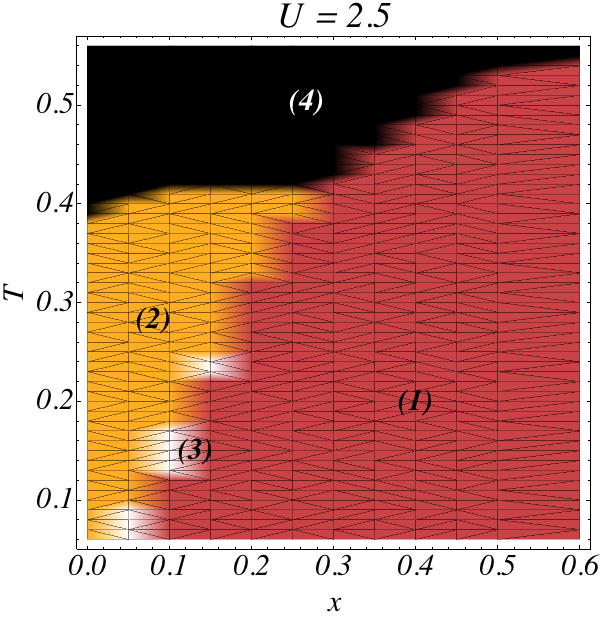}  & \includegraphics[width=5cm]{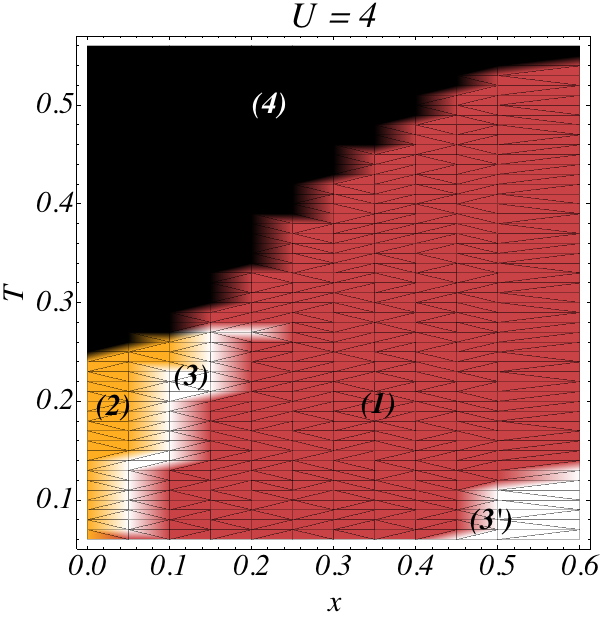}  & \includegraphics[width=5cm]{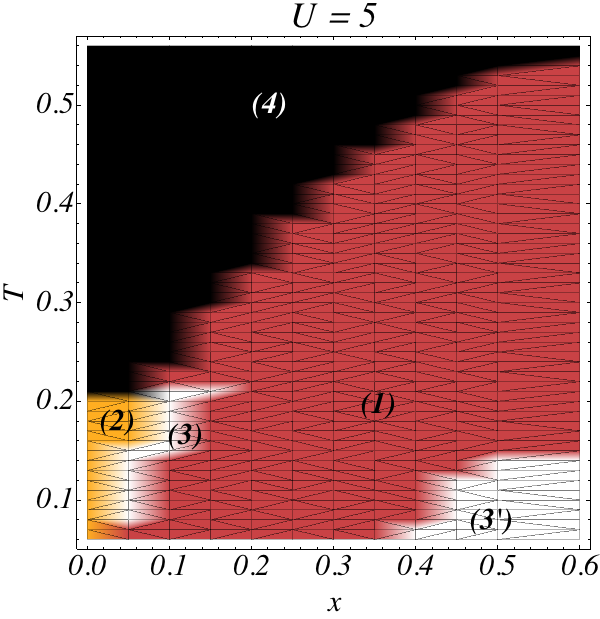} \tabularnewline
\end{tabular}\caption{\label{fig:1D_PD}Mean Field phase diagram in the $x-T$ plane as
a function of $U$ for a chain with $200$ sites. The colors represent
the different types of solution minimizing the free energy: (1) $\chi\neq0,\,\Delta=0$
(Red); (2) $\chi=0,\,\Delta\neq0$ (Orange); (3) $\chi\neq0,\,\Delta\neq0$
(White); (4) $\chi,\Delta=0$ (Black). The calculated points are placed
in the nodes of the finite mesh of the values of $x$ and $T$ considered,
the colors plotted between nodes are interpolated. }

\end{figure*}

The above treatment should be valid for arbitrary values of the interaction
parameter $U$ once the double occupancy described by the $d$ bosonic
field is fully taken into account. This should be of great importance
near half-filling because in this regime $\av{d^{\dagger}d}$ is not
small compared to $\av{e^{\dagger}e}$.

\begin{figure}
\includegraphics[width=5cm]{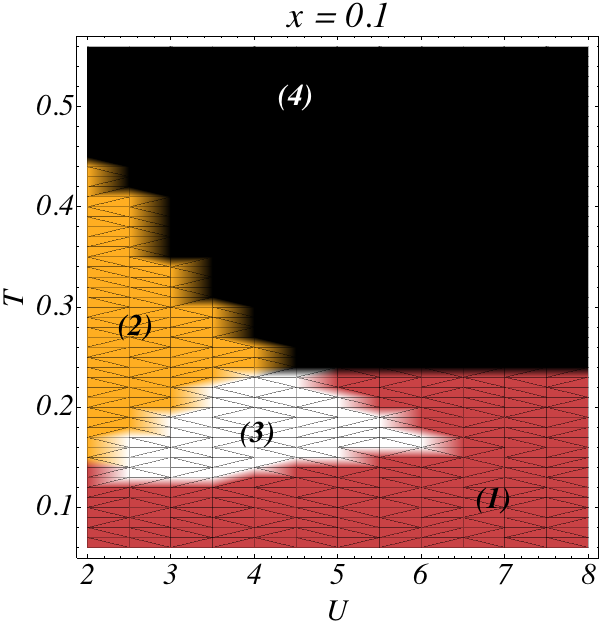}\caption{\label{fig:1D_T}Mean Field phase diagram in the $U-T$ plane for
$x=0.1$ for a chain with $200$ sites. The colors represent the different
types of solution minimizing the free energy as in Fig. \ref{fig:1D_PD} }

\end{figure}

Contrary to the $t-J$ model no canonical transformation was performed
at this stage in the physical electrons. Note that discarding the
$d$ field at this stage would yield diagonal $\mathbf{F}$ and $\mathbf{B}$
matrices and the subsequent MF decoupling would miss the phases of
the ZA fermions with non-zero pairings $\Delta$. This kind of phases
could however be considered if one adopts a variational procedure
(Ref. \cite{Brinckmann_2001}). The $t-J$ model also presents an
$\mathbf{S}\cdot\mathbf{S}$ term not present in Eq. (\ref{eq:H-S-Lag})
which is responsible for sub-lattice magnetization in the ground state
of the square lattice Hubbard model near half-filing. Anti-ferromagnetic
correlations will nevertheless be generated if one integrates out
the $d$ fields before doing the MF decoupling; however, such procedure
leads to six-body coupling terms and will not be considered here.
Even if Eq. (\ref{eq:H-S-Lag}) can not produce sub-lattice magnetization
we expect antiferromagnetic spin-spin correlation functions for non-frustrated
lattices. In this work no boson condensation is considered away from
$T=0$; this corresponds to a fully 1D or 2D model of holons $e$
and doublons $d$ \cite{Lee_2006}.

Eqs. (\ref{eq:MF-1}-\ref{eq:MF-3}) together with the fixed doping
condition were solved numerically in one and two dimensions for several
values of $U/t$ (typically in the range of 2 to 6) , for different
values of doping $x\simeq0,...,0.6$ and temperature. For the 2D case
no particular symmetry based ansatz is implemented leading to some
peculiar phases. Before discussing some of the MF solutions in detail
a few remarks are in order:

\noindent (i) Some care was taken diagonalizing the fermionic and
bosonic quadratic Hamiltonians since the hoppings and pairings found
were in general complex numbers. However some of the converged solutions
lead to non-Hermitian Hamiltonians. These solutions were discarded
as {}``unphysical''. However, for the slave-particle approach the
only physical constraints are for the composite electron operators
and so maybe some of these solutions could be physically meaningful.

\noindent (ii) The convergence of the solutions was quite difficult
for some regions, specially near zero doping and for very low temperatures.
For two dimensions a few points in the $U-x-T$ space were tested
against finite size effects running the calculations in a $32\times32$
lattice (instead of $16\times16$) with only small quantitative changes
in the results. However near regions of phase competition small finite
size effects can change the global minimum yielding qualitatively
different results.

\begin{figure*}[!t]
 \begin{tabular}{lll}
\begin{tabular}{c}
\includegraphics[width=5cm]{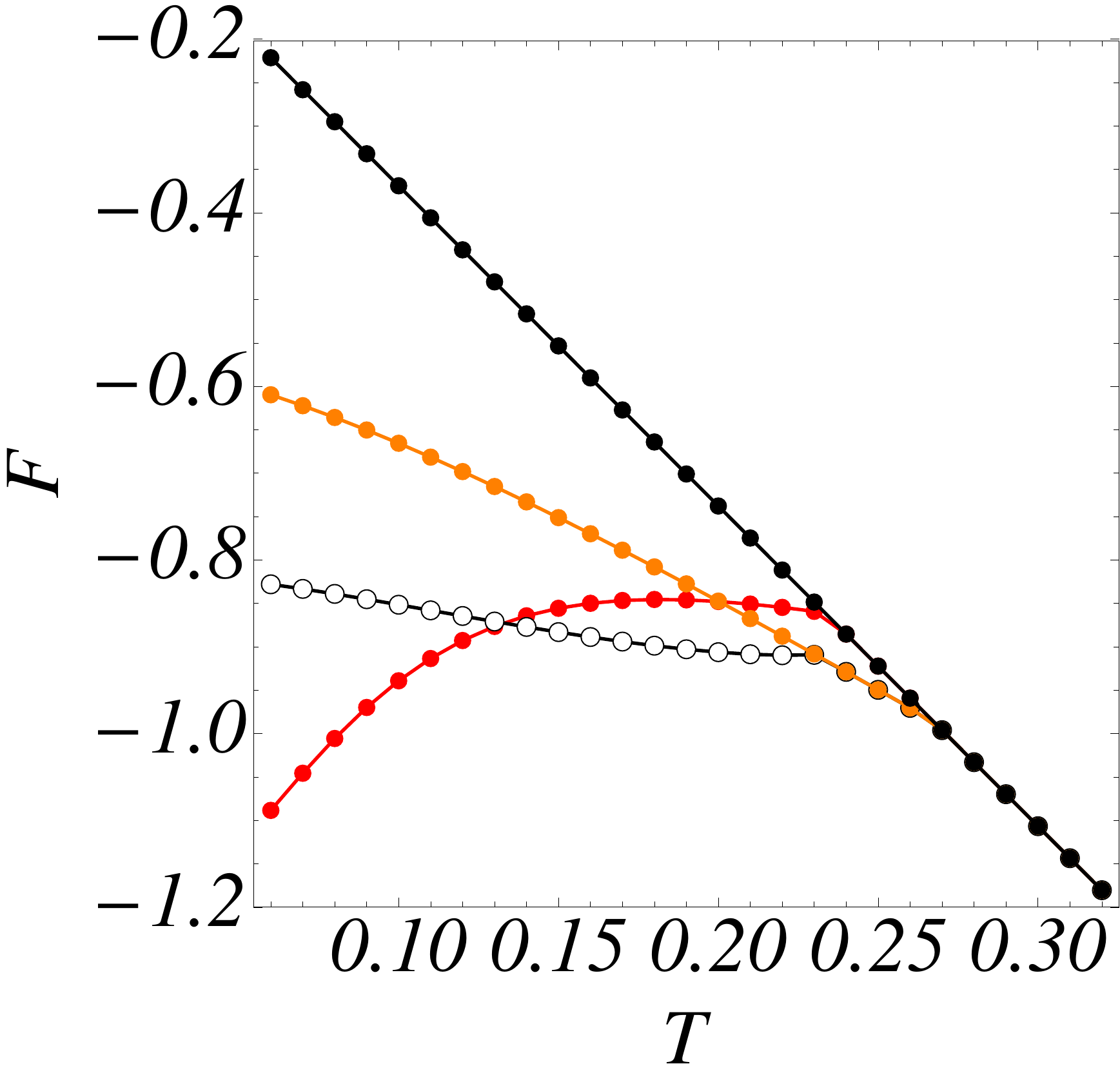}\tabularnewline
\end{tabular} & \begin{tabular}{cc}
\includegraphics[width=3cm]{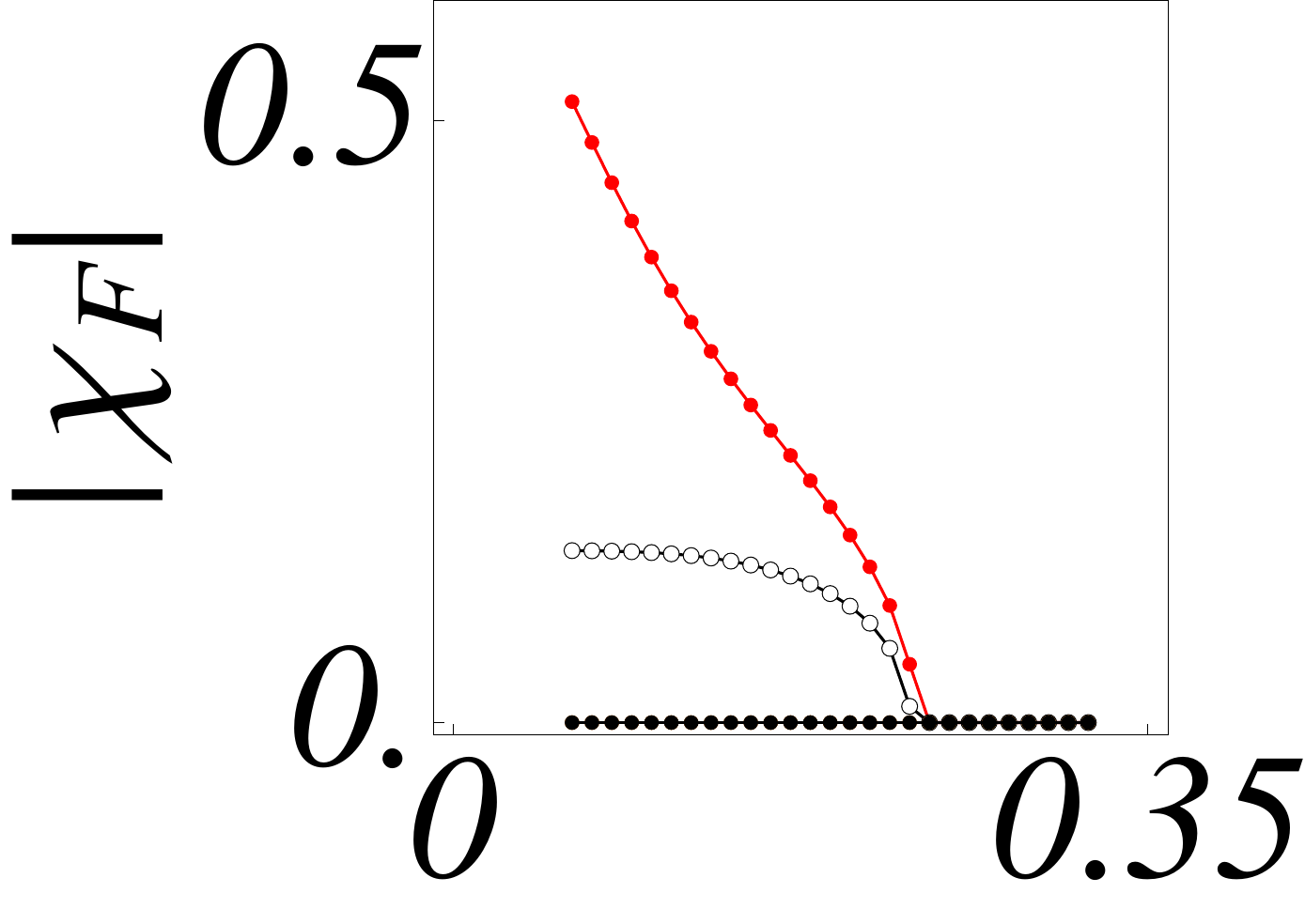}  & \includegraphics[width=3cm]{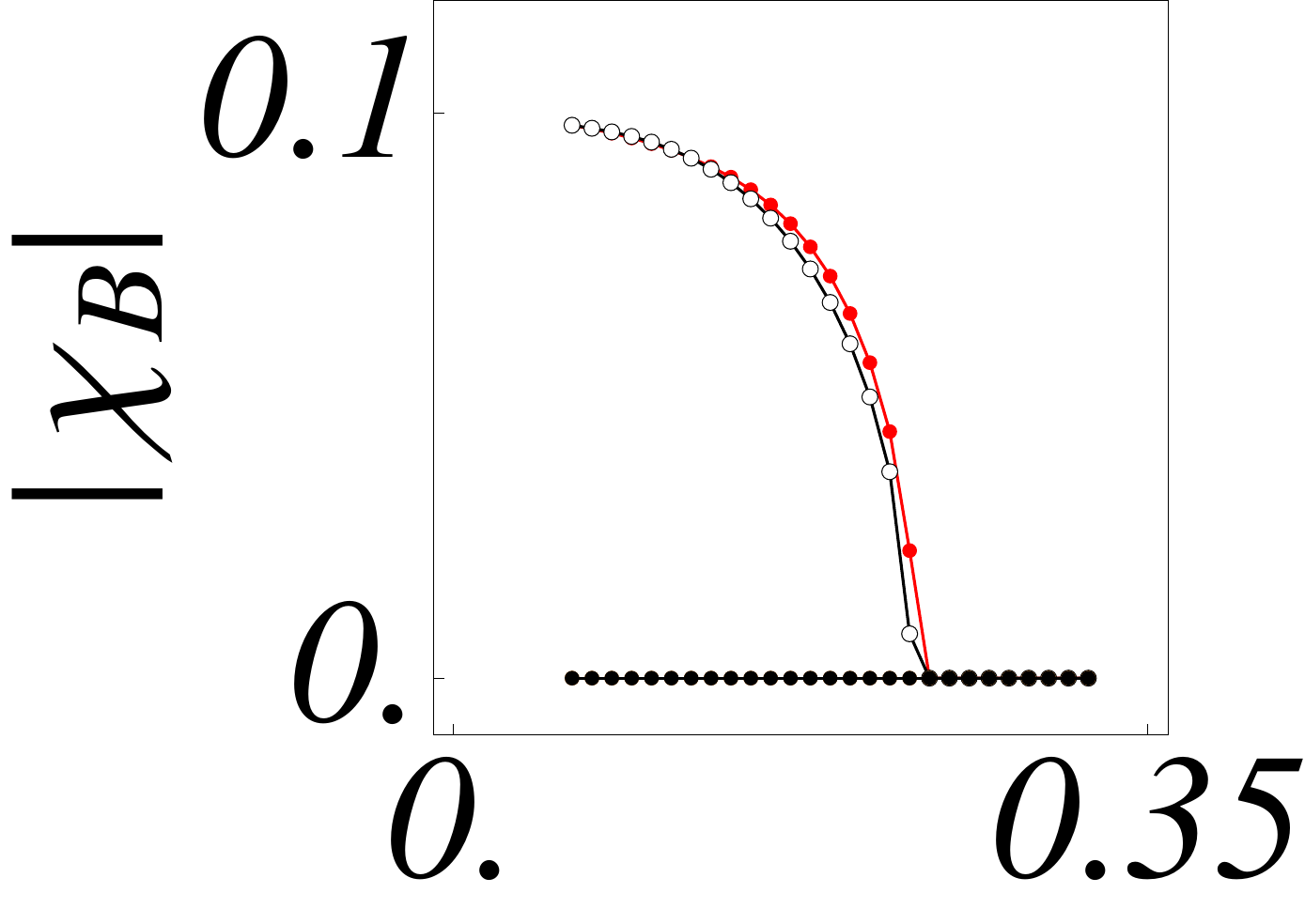}\tabularnewline
\includegraphics[clip,width=3cm]{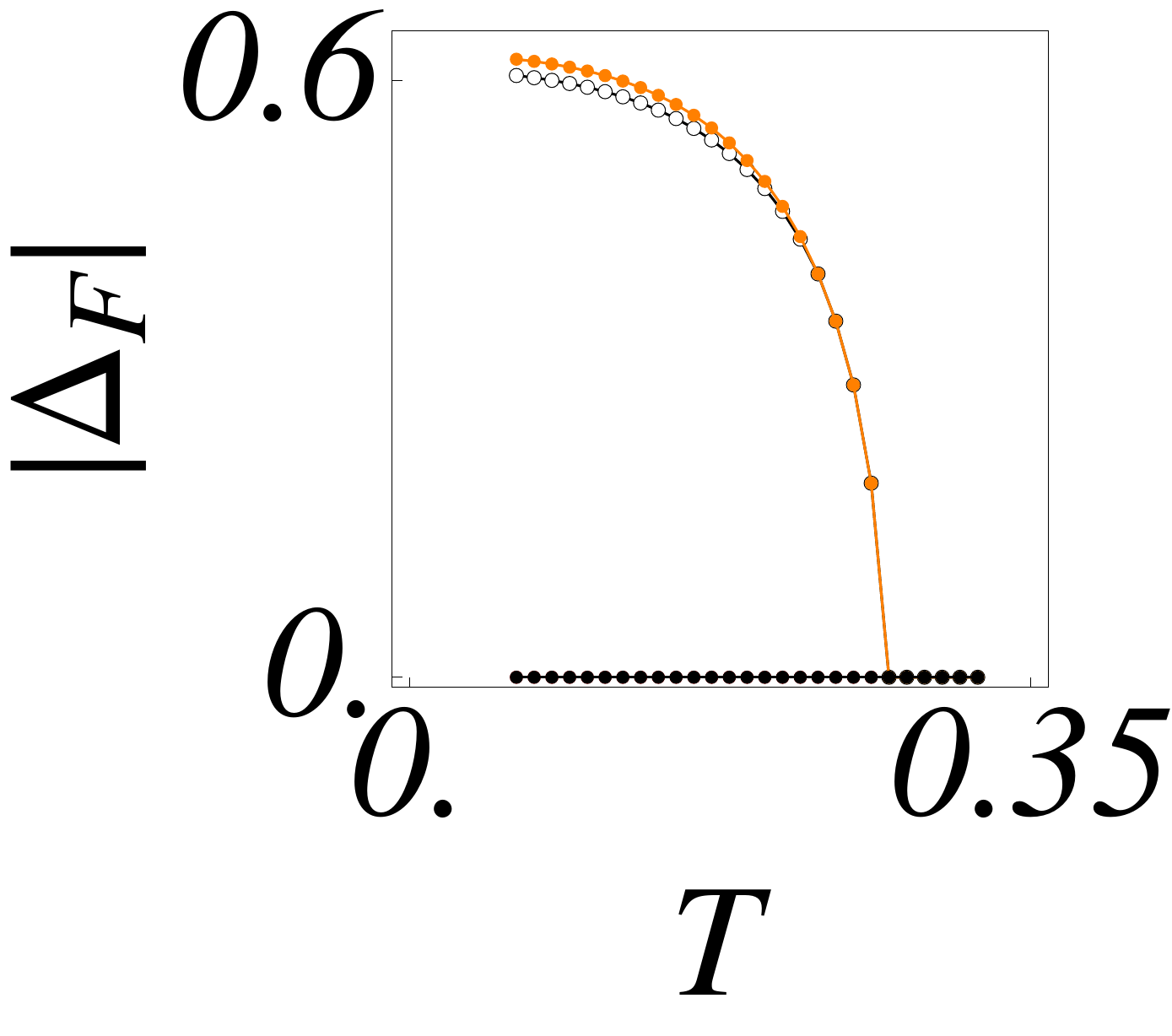}  & \includegraphics[clip,width=3cm]{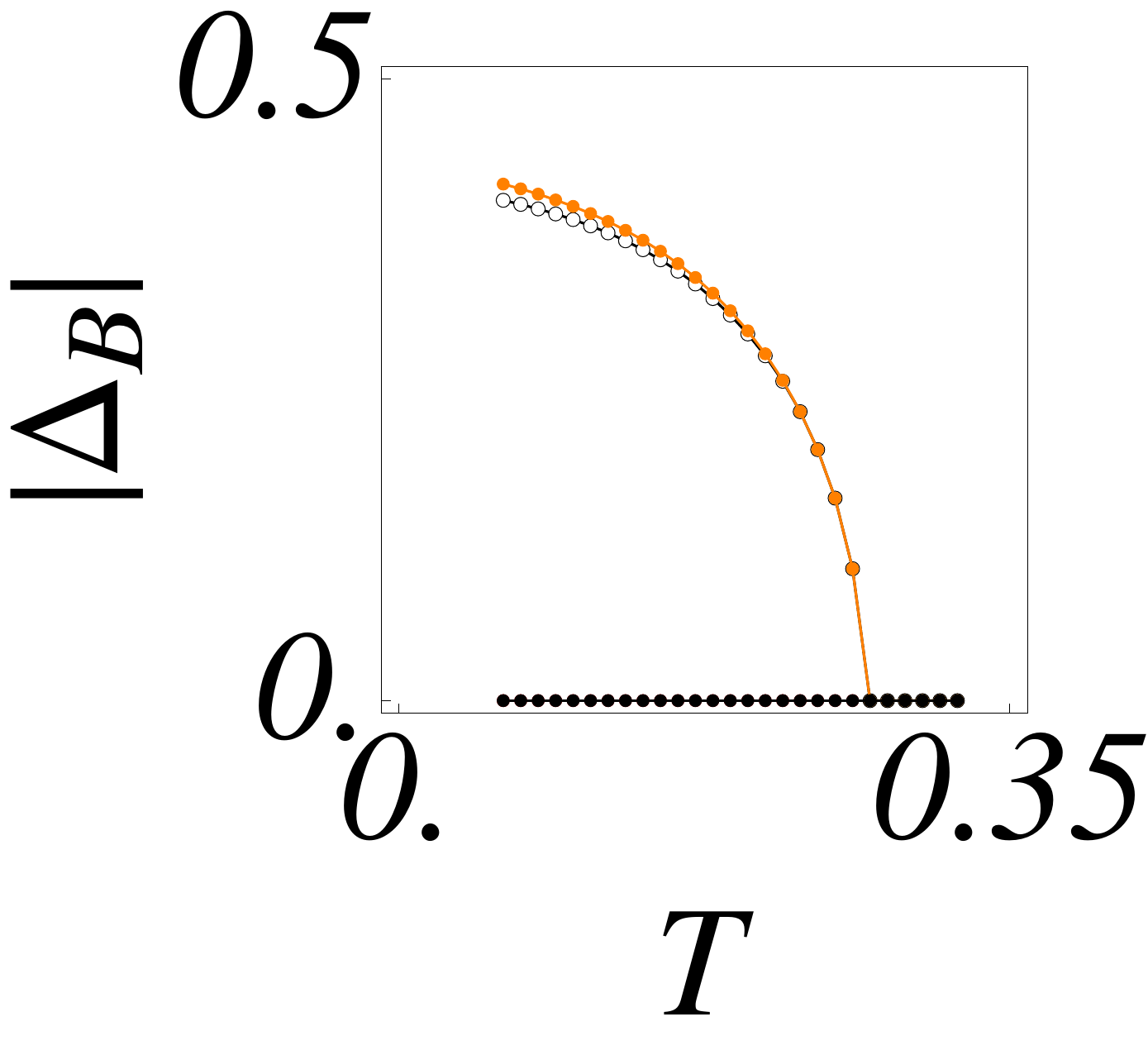}\tabularnewline
\end{tabular} & \begin{tabular}{c}
\includegraphics[width=5cm]{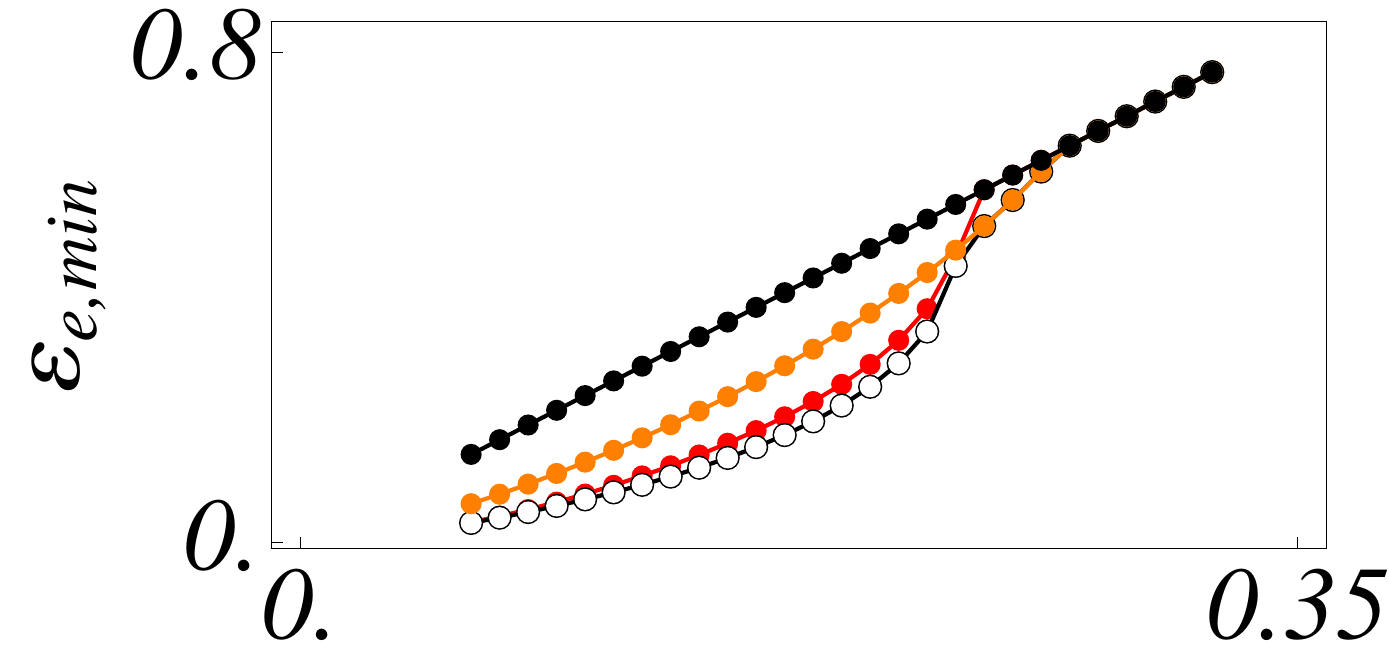}\tabularnewline
\includegraphics[width=5cm]{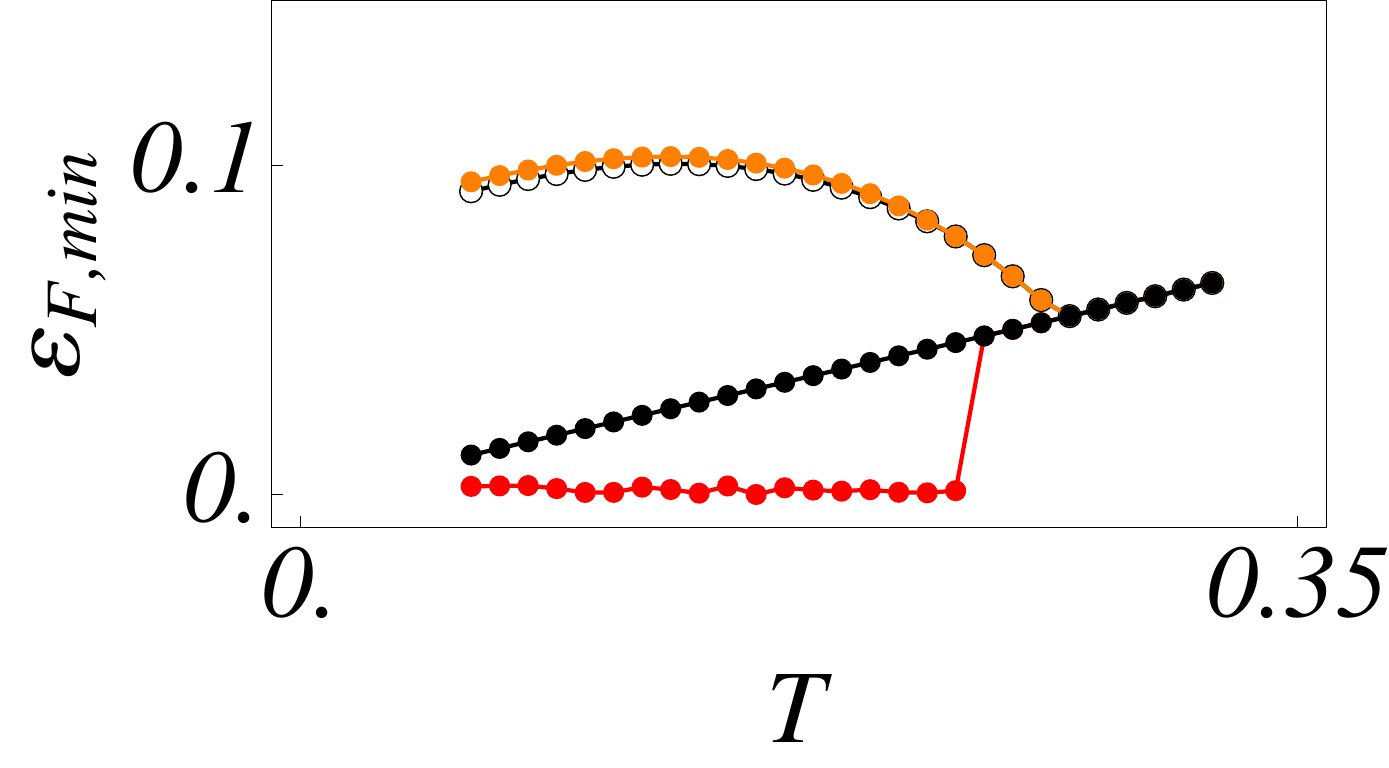}\tabularnewline
\end{tabular}\tabularnewline
\end{tabular}\caption{\label{fig:1D_FreeEn_T}Left panel: Free energy for 1D, $U=4,x=0.1$
as a function of the temperature. The Black, White and Red dots follow
the same color code as the phase diagrams. Central panel: Fermionic
and bosonic order parameters as a function of temperature for each
phase separately. Right Panel: Minimum energy as a function of temperature
for the fermionic and bosonic bands.}

\end{figure*}
\noindent (iii) For real physical systems described approximately
by the Hubbard Hamiltonian other small coupling terms are expected
that can qualitatively change the phase diagram locally. Different
stable solutions that are not global minima will be considered elsewhere
and are listed in the Appendix. %

\section{Mean Field Phase Diagram for Hubbard chain and square lattice}

A MF treatment of the original electron problem is not expected to
be a good approximation for the one-dimensional case. However, it
is interesting to construct a MF treatment in terms of new operators
and to test the differences with the existing exact results and the
domain of validity of the present treatment. It also permits to have
a generic idea of the phase diagram and of the different phases present
in higher dimensions.

In the case of the 1d Hubbard model we present some of the MF solutions
obtained numerically solving the MF Eqs. (\ref{eq:MF-1}-\ref{eq:MF-3})
on a $200$ site chain. The results were obtained as follows: for
a chosen point of the parameter space $U-T-x$ we generated several
random trial solutions and used them as a starting point to our numerical
routine. We obtained several different solutions unrelated by gauge
transformations; however, just 4 of these solutions are relevant for
the range of parameters considered here. The other solutions have
very high values of the free energy or are {}``unphysical''. These
4 solutions were extended to the rest of the $U-x-T$ parameter space
using as initial conditions a nearby converged solution. The four
physical MF solutions differ by the existence of non-zero hopping
and pairing terms. Fig. \ref{fig:1D_PD} shows the phase diagram $x-T$
for different values of $U$ and Fig. \ref{fig:1D_T} the phase diagram
$U-T$ for $x=0.1$. In these figures the results are presented on
a finite mesh of points in the $x-T$ or $U-T$ planes.

For the two dimensional case we present the most stable MF solutions
obtained numerically by solving the MF Eqs. \ref{eq:MF-1}-\ref{eq:MF-3}
on a $16\times16$ square lattice. The results were obtained as follows:
for a chosen point of the parameter space $U-T-x$ we generated 1000
random initial conditions and used them as a starting point to our
numerical routine. From this 1000 initial conditions 30 different
solutions, non gauge equivalent, were found to converge. After this
first step the 30 different solutions were extended to the rest of
the $U-x-T$ parameter space using as initial conditions a nearby
converged solution. This procedure is tedious since the convergence
of the solutions is not always easy and some of them do not converge
even if the difference in the parameters is small. Finally for each
point in the $U-x-T$ space (again defined on a finite mesh) the solution
with smallest free energy was found. Note that we did not impose any
particular symmetry to solve the MF equations, the only assumption
being translational invariance ($\boldsymbol{Q}_{r,\delta}=\boldsymbol{Q}_{\delta}$),
in order to diagonalize the system in momentum space. That fact explains
the proliferation of solutions of the MF equations.

\subsection{Description of the Phase Diagram }

Generically the phases found solving the MF solutions are characterized
as follows:

- Phase (1) Conducting phase characterized by $\chi\neq0,\ \Delta=0$
(Red): the spins are gapless and the charge degrees of freedom present
a gap of the order of the temperature, which closes at $T=0$.

- Phase (2) $\chi=0,\ \Delta\neq0$ phase (Orange), gapped for both
degrees of freedom. Since it appears near $x=0$ it is tempting to
identify this phase as an insulating antiferromagnet.

- Phase (3) $\chi\neq0,\ \Delta\neq0$ phase (White): precursor of
the superconductor, in this phase there exists spin singlet formation
but the charge motion is incoherent since no condensation was allowed.
If we had imposed $\av{e_{k=0}}=Z$ this phase would split in two
sub-phases analog to the pseudogap and superconducting phases in \cite{Lee_2006}.

- Phase (4) High energy phase (Black): is an incoherent phase where
all correlations are zero.

\subsection{Characterization of the Phases}

Our numerical results in one and two dimensions show that the $\chi=0,\ \Delta\neq0$
(Orange) phase is dominant in the low doping region up to a temperature
that decreases with U. In the 1D case this phase extends to the under-doped
region and interfaces with the $\chi\neq0,\ \Delta=0$ (Red) phase,
present at higher doping, by a small finite-energy region where $\chi\neq0,\ \Delta\neq0$
(White phase). In 2D the $\chi=0,\ \Delta\neq0$ (Orange) phase is
numerically unstable and we could only find it for zero doping. The
$\chi\neq0,\ \Delta\neq0$ phase appears also in the high doping regime
at low $T$. In 1D the size of this region grows clearly with U, however,
in 2D this is not clear but is definitely present at low energy. At
very low $T$ the most stable solution is the Red phase ($\chi\neq0;\Delta=0$)
except at half-filling.\textit{}\\
\textit{}\\
\textit{One-dimensional case}

Figure \ref{fig:1D_FreeEn_T} shows the free energy of the Hubbard
chain for the four MF solutions when $U=4,x=0.1$ as a function of
the temperature. At low temperature the $\chi\neq0$ (Red) phase has
a lower free energy and there is a first order phase transition with
increasing temperature as the $\chi\neq0;\Delta\neq0$ phase becomes
less energetic. At higher temperatures two second order phase transitions
occur: first, for $T\simeq0.23$, $\chi$ decreases to zero as the
$\chi\neq0;\Delta\neq0$ joins the $\chi=0;\Delta\neq0$ solutions;
the second phase transition occurs for $T\simeq0.27$ when $\Delta$
vanishes (see Fig. \ref{fig:1D_FreeEn_T}- central panel).

In the right panel of the same figure we show the minimum energy of
the various bands as a function of temperature for the various phases.
At low temperature where the Red phase is the most stable the fermionic
bands are gapless. This phase is always gapless up to the point where
it merges with the fully incoherent (black) phase. In the other phases
the fermions are gapped (note that the spin spectrum in the Orange
and White phases is almost the same). On the other hand, the bosonic
(charge) spectrum is always gapped except at zero temperature.

Considering for instance $U=4,T=0.01$ we may as well analyze the
results as a function of doping. At zero doping both $\chi\neq0;\Delta\neq0$
and $\chi=0;\Delta\neq0$ coincide, this degeneracy is lifted for
finite doping (in a very narrow region) and the $\chi\neq0:\Delta\neq0$
phase presents the minimal value of the free energy. From $x\simeq0.1$
to $x\simeq0.5$ there is an intermediate phase $\chi\neq0;\Delta=0$
delimited by two first order transitions. For higher doping the $\chi\neq0;\Delta\neq0$
regains the minimal value of the free energy.

\begin{figure}
\begin{tabular}{ll}
\includegraphics[width=4cm]{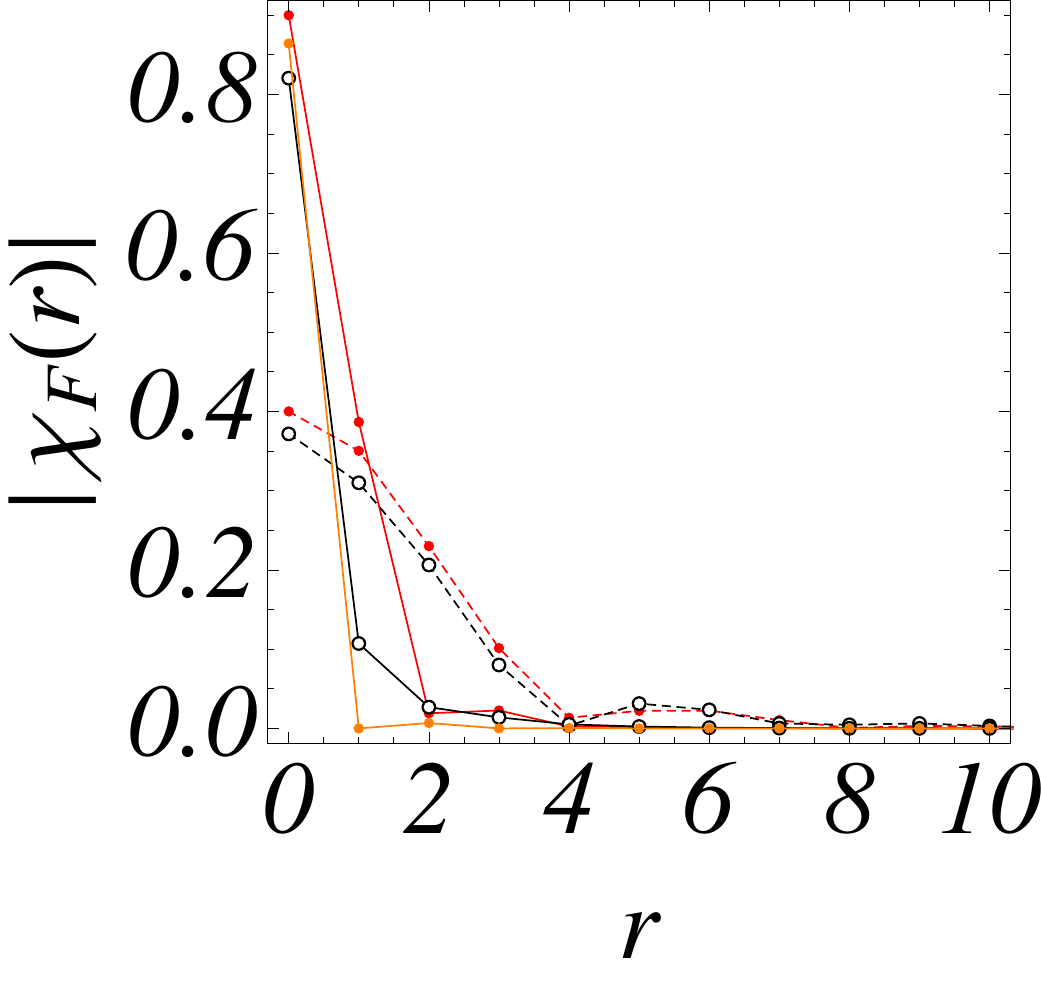}  & \includegraphics[width=4cm]{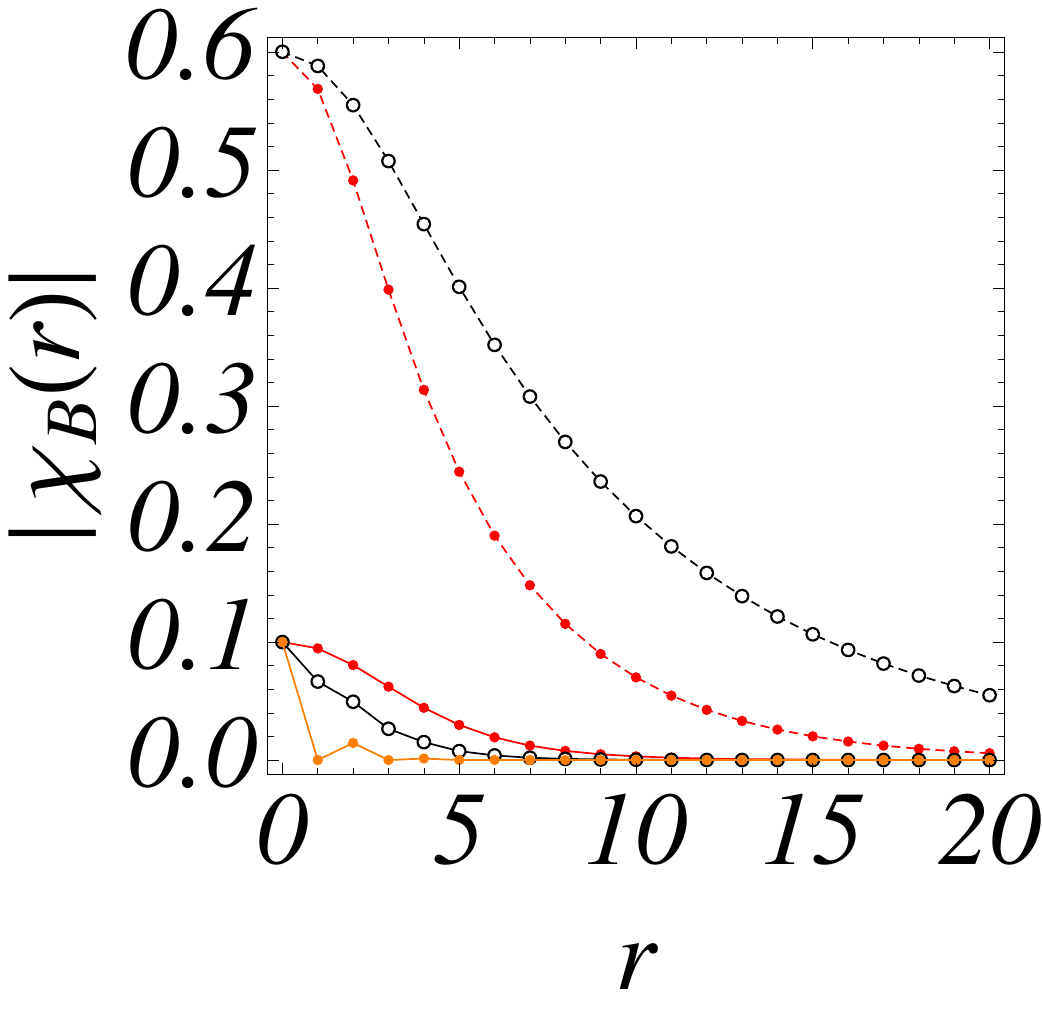}\tabularnewline
\includegraphics[width=4cm]{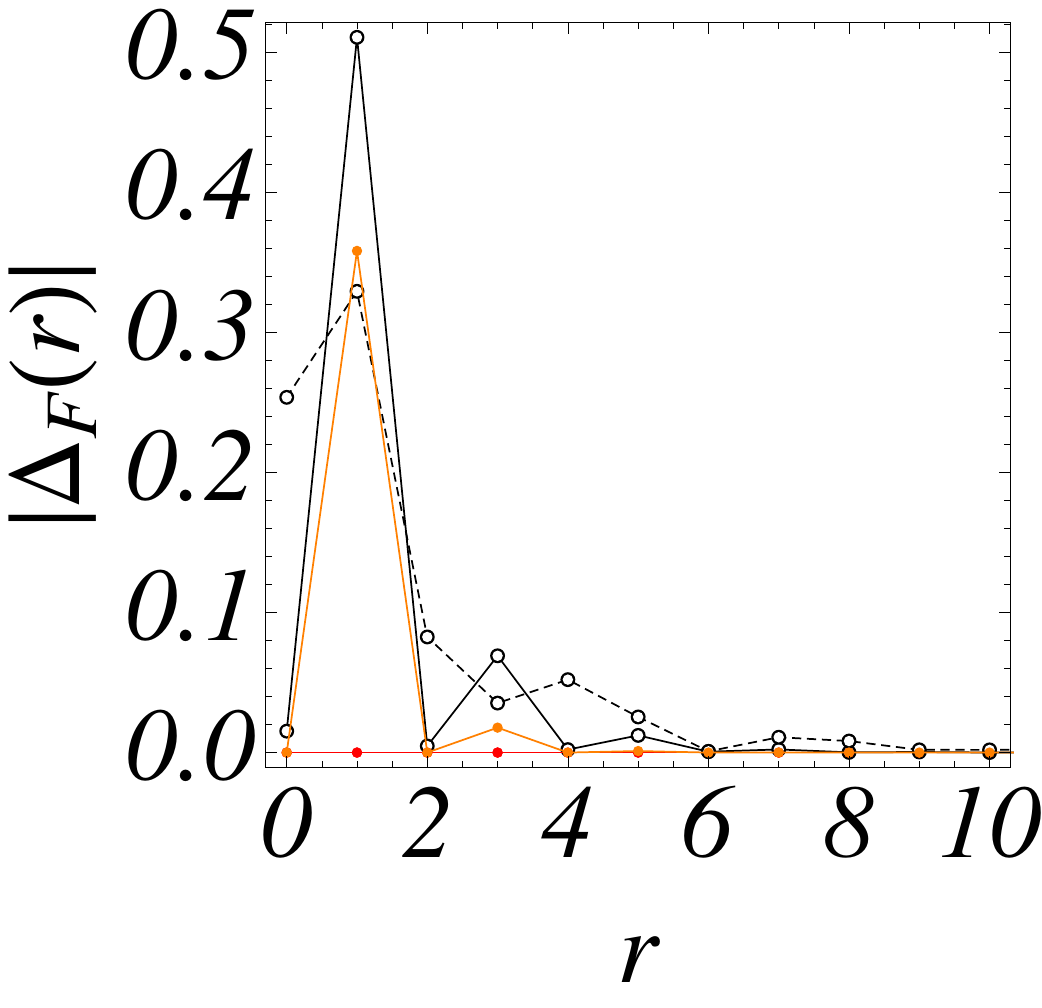}  & \includegraphics[width=4cm]{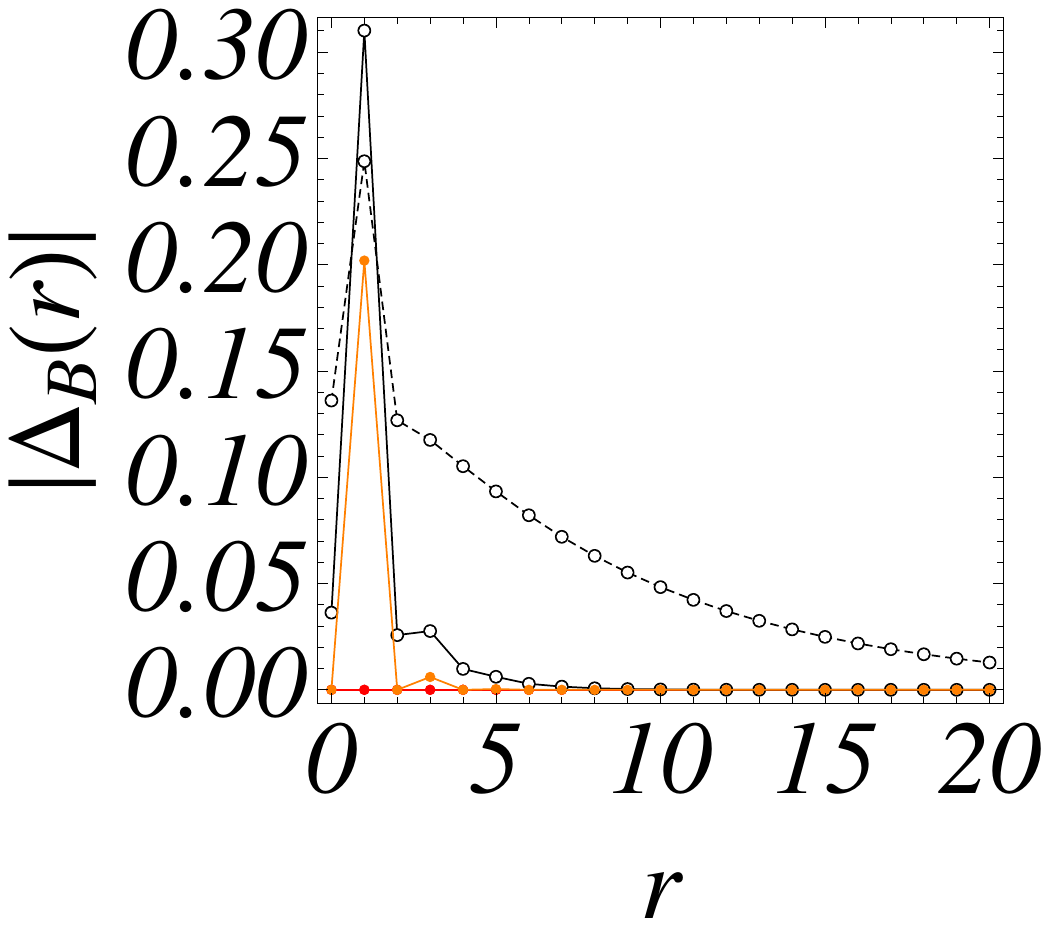}\tabularnewline
\end{tabular}\caption{\label{fig:1D_NonPhys}Hopping and pairing correlations as a function
of the distance in 1D, computed for the three non-trivial MF phases
with $U=4$. Red(full): Phase $\chi\neq0,\Delta=0$ computed for $T=0.10,x=0.1.$
Red(dashed): Phase $\chi\neq0,\Delta=0$ computed for $T=0.15,x=0.6$.
White: Phase $\chi\neq0,\Delta\neq0$ computed for $T=0.19,x=0.1.$
White (dashed) : Phase $\chi\neq0,\Delta\neq0$ computed for $T=0.08,x=0.6.$
Orange: Phase $\chi=0,\Delta\neq0$ computed for $T=0.24,x=0.1.$}

\end{figure}

The hopping and pairing correlation functions 
\begin{eqnarray}
\chi_{F}(r) & = & \av{\mathit{s}_{r,1}^{\dagger}\mathit{s}_{0,1}+\mathit{s}_{r,-1}^{\dagger}s_{0,-1}}_{0},\nonumber \\
\Delta_{F}(r) & = & \av{\mathit{s}_{r,1}\mathit{s}_{0,-1}-\mathit{s}_{r,-1}\mathit{s}_{0,1}}_{0},\nonumber \\
\chi_{B}(r) & = & \av{\mathit{d}_{r}^{\dagger}\mathit{d}_{0}-\mathit{e}_{r}^{\dagger}\mathit{e}_{0}}_{0},\nonumber \\
\Delta_{B}(r) & = & \av{\mathit{d}_{r}\mathit{e}_{0}+\mathit{e}_{r}\mathit{d}_{0}}_{0},\label{eq:non_phys_corr}
\end{eqnarray}
are shown in Fig. \ref{fig:1D_NonPhys} at two doping values $x=0.1,x=0.6$.
Even if these correlation functions are not gauge invariant they can
be quite useful to characterize the different phases. In particular
one can clearly see the difference between the two disjoint $\chi\neq0;\Delta\neq0$
(White) ((3) and (3')) regions. At lower doping $x=0.1$ we consider
the three non-trivial solutions as a function of increasing temperature
and at the higher doping ($x=0.6$) we consider the White and Red
solutions. Both for the fermion and the boson hopping correlation
functions the correlation length increases as the doping increases.
Particularly the bosonic correlation function has a large correlation
length. Analyzing the correlation length of $\Delta_{B}$ one clearly
sees a long range correlation in the high doping regime possibly precursor
of Bose-condensation and superconductivity. In the low doping region
both the bosonic and the fermionic correlation functions have a smaller
range consistent with a spin gapped state. In this regime the two
correlation functions have similar range while at higher doping the
charge correlation function has a much larger range compared to the
spin correlation function.\\
\begin{figure*}
\begin{tabular}{ccc}
\includegraphics[width=5cm]{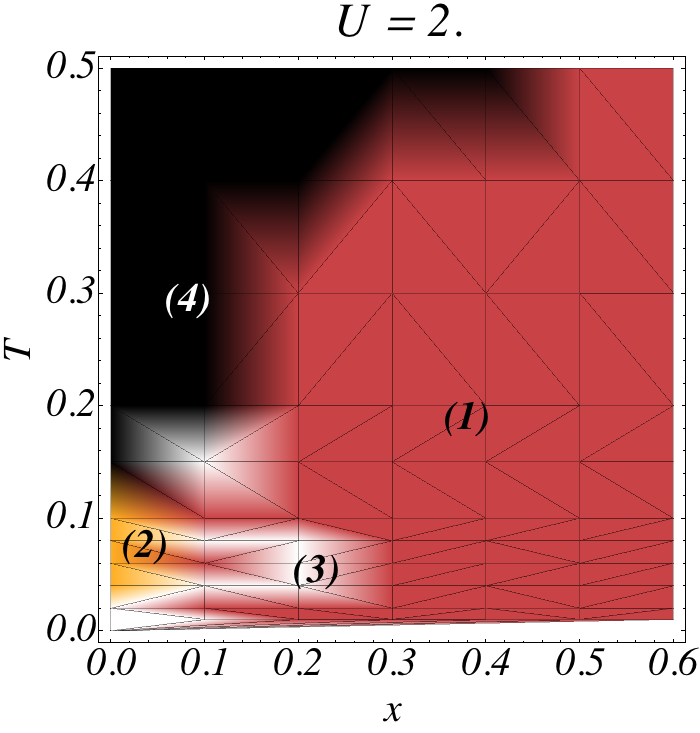}  & \includegraphics[width=5cm]{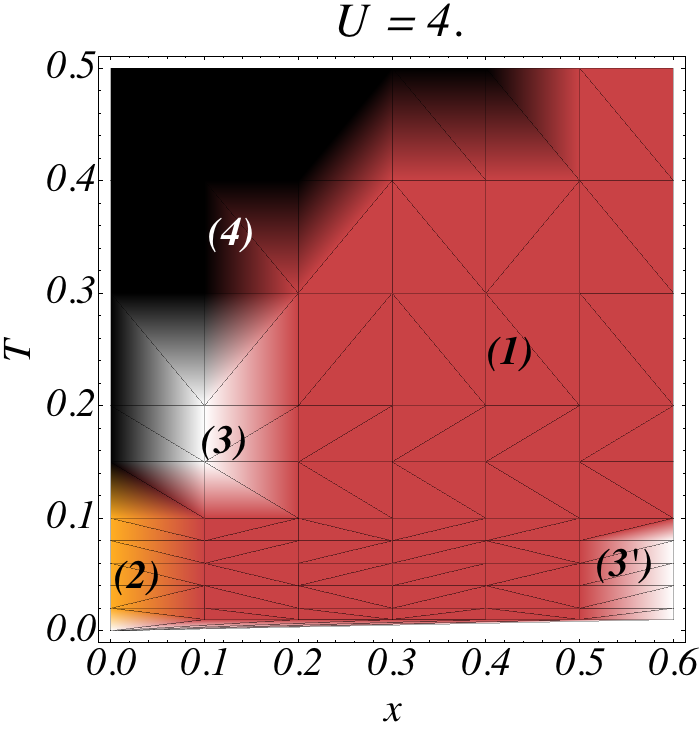}  & \includegraphics[width=5cm]{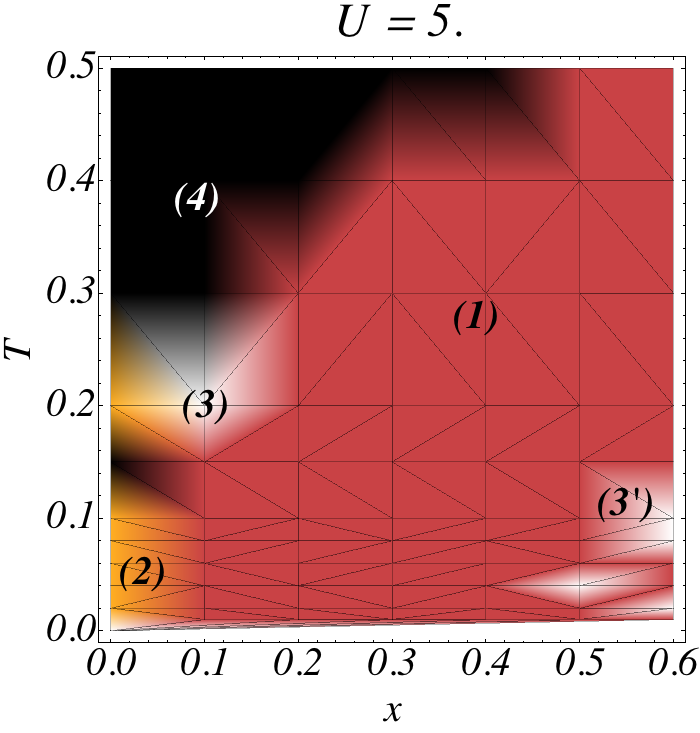} \tabularnewline
\end{tabular}\caption{\label{fig:2D_PD-1}Mean field phase diagram in the $x-T$ plane as
a function of $U$ for a $16\times16$ square lattice. The colors
represent the different types of solution minimizing the free energy:
(1) $\chi\neq0,\,\Delta=0$ (Red); (2) $\chi=0,\,\Delta\neq0$ (Orange);
(3) $\chi\neq0,\,\Delta\neq0$ (White); $\chi,\Delta=0$ (Black).
The calculated points are placed in the nodes of the mesh, the colors
plotted between nodes are interpolated. }

\end{figure*}
\begin{figure*}[!t]
 \begin{tabular}{lll}
\begin{tabular}{c}
\includegraphics[height=5cm]{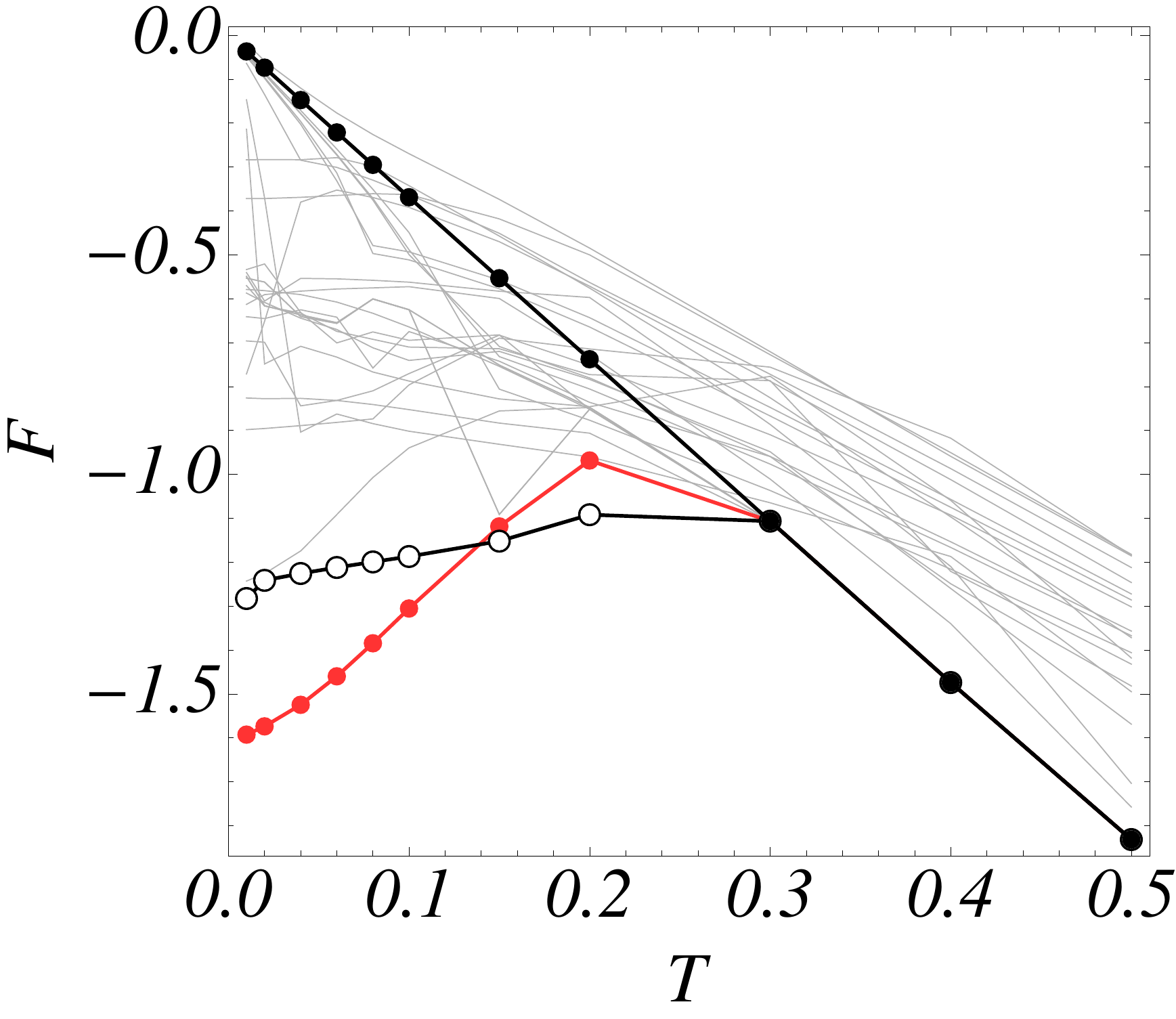}\tabularnewline
\end{tabular} & \begin{tabular}{cc}
\includegraphics[width=3cm]{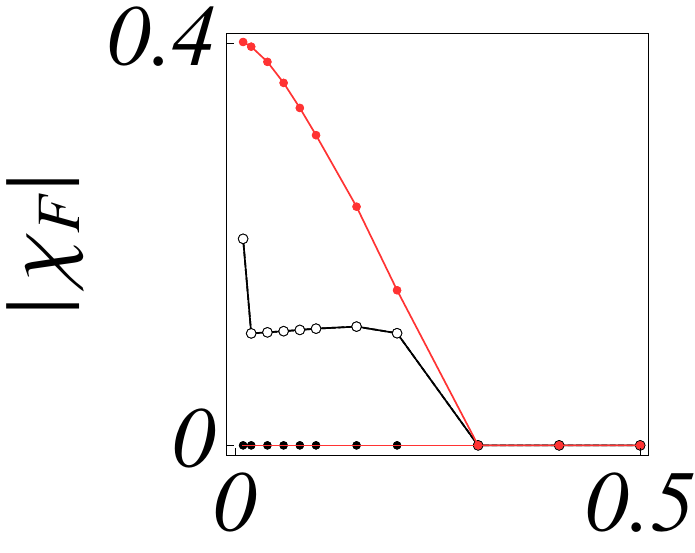}  & \includegraphics[width=3cm]{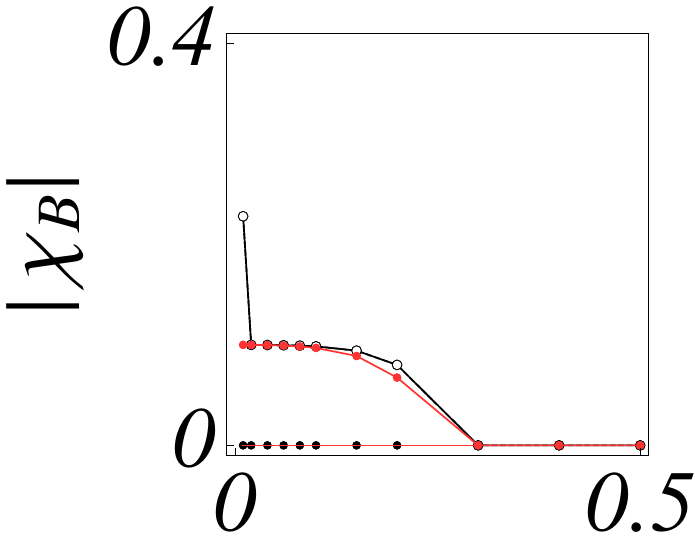}\tabularnewline
\includegraphics[clip,width=3cm]{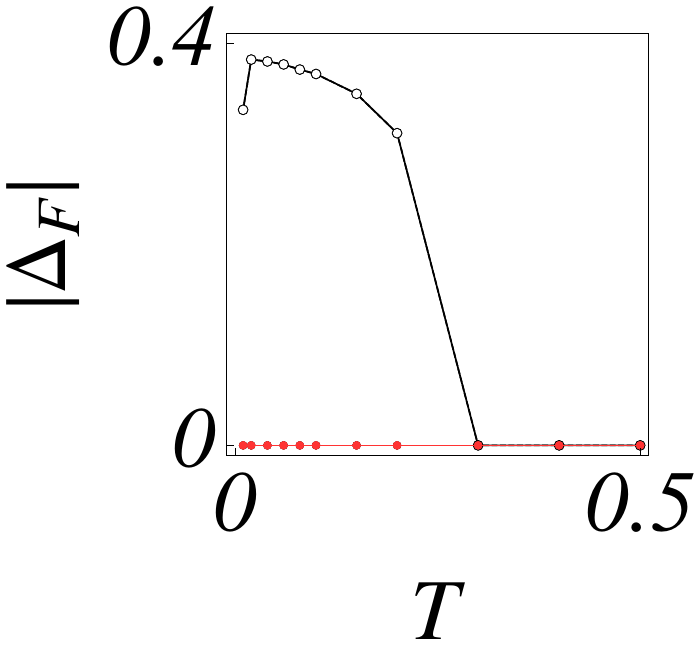}  & \includegraphics[clip,width=3cm]{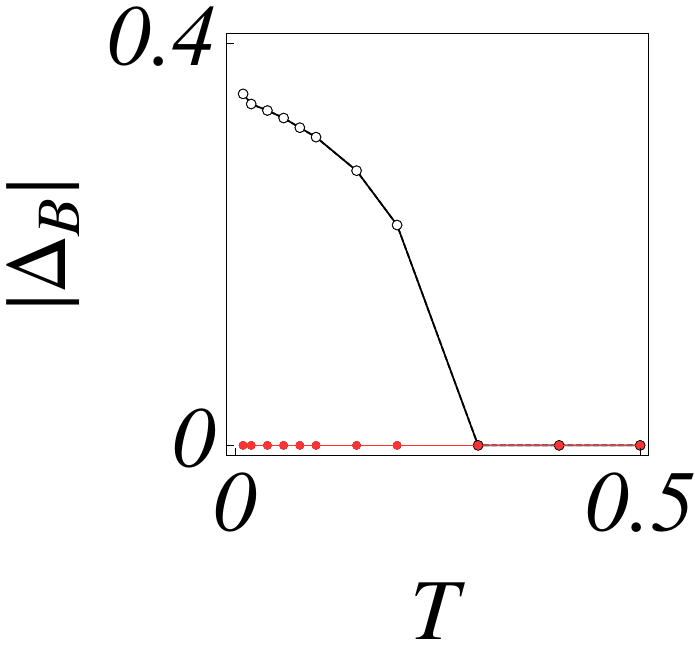}\tabularnewline
\end{tabular} & \begin{tabular}{c}
\includegraphics[width=5cm]{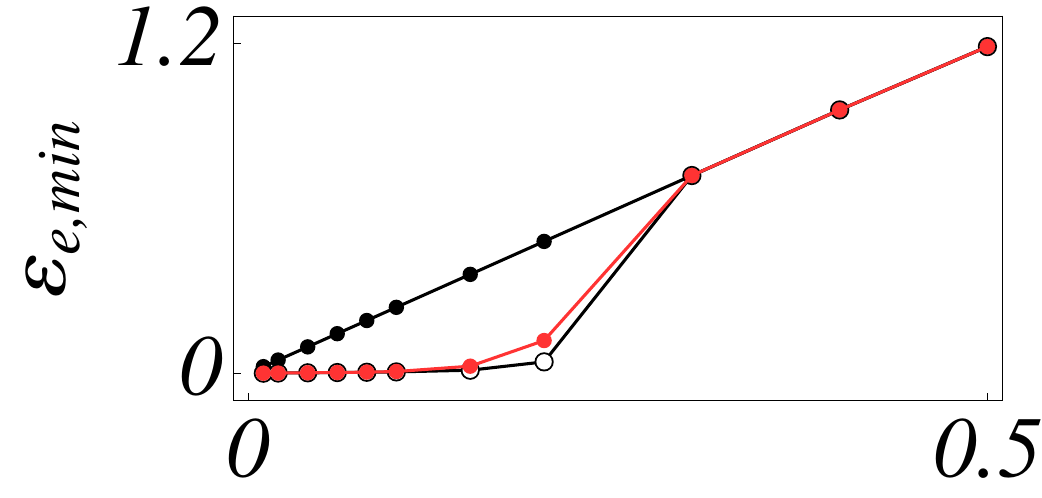}\tabularnewline
\includegraphics[width=5cm]{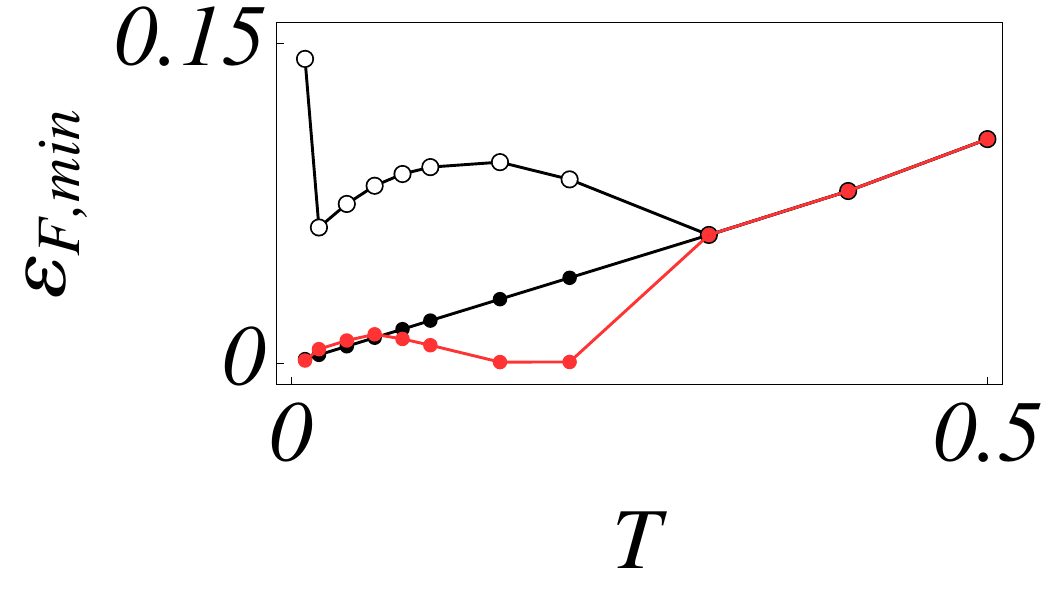}\tabularnewline
\end{tabular}\tabularnewline
\end{tabular}\caption{\label{fig:2D_FreeEn_T}Left Panel: Free energy for $U=4,x=0.1$ as
a function of the temperature. The Black, White and Red dots follow
the same color code as the phase diagrams. The Gray lines in background
are the free energies of the rest of the converged solutions (most
of which are unphysical, see main text). Central panel: Fermionic
order parameters as a function of temperature for each phase separately,
note that solutions that minimize the free energy are symmetric under
rotations of the lattice. Right Panel: Minimum of the one particle
excitation energies of the diagonalized bosonic and fermionic MF Hamiltonians.}

\end{figure*}
\noindent \textit{Two-dimensional case}

\noindent The phase diagram for the square lattice is shown in Fig.
\ref{fig:2D_PD-1}. The free energy for the square lattice as a function
of the temperature is shown in Fig. \ref{fig:2D_FreeEn_T} for $U=4,\ x=0.1$.
The low temperature state is, as in the 1D case, given by the phase
$\chi\neq0,\Delta=0$ (Red). There is a first order phase transition
to the solution $\chi\neq0;\Delta\neq0$ (White) as the temperature
increases. However contrary to the 1D case there is only one second
order phase transition for higher temperature where both hoppings
and pairings vanish at the same time. Both phases (Red and White) present gapless
bosonic (charge) excitations (see Fig. \ref{fig:2D_FreeEn_T}-Right
panel) but the minimum of the bosonic 1-particle excitations is located
at $k=\left\{ 0,0\right\} $ and $k=\left\{ \pi,\pi\right\} $ respectively.
Furthermore the $\chi\neq0,\Delta=0$ (Red) phase has no spin gap
(see Fig. \ref{fig:2D_FreeEn_T}-Right panel, here we believe that
the oscillations are due to finite size effects). Also note that contrarily
to the one-dimensional case the fully incoherent phase has a fermionic
(spin) gap smaller than the White phase. However, the White phase
is the most stable one for dopings between $0.15-0.3$ which indicates
the presence of a spin gap (pseudogap) in this region.

As a function of doping the free energy is minimal at half filling
for the state $\chi=0;\Delta\neq0$ (Orange). This state is the zero
doping limit of the state $\chi\neq0;\Delta\neq0$ (White) where the
values of $\chi$ vanish. Moreover one observes a first order phase
transition to the state $\chi\neq0,\Delta=0$ (Red) as the doping
is increased. For $U=2$ this arises away from half-filling resulting
in two consecutive phase transitions with increasing values of $x$,
the first from $\chi=0;\Delta\neq0$ (Orange) to $\chi\neq0;\Delta\neq0$
(White) is of second order, and the second from $\chi\neq0;\Delta\neq0$
(White) to$\chi\neq0,\Delta=0$ (Red) is of first order. For $U=4,5$
only the first order transition is observed arising at $x=0$ between
phases $\chi=0;\Delta\neq0$ (Orange) and $\chi\neq0,\Delta=0$ (Red).
For still higher doping and sufficiently strong $U$ there is another
first order transition back to the state $\chi\neq0;\Delta\neq0$
(White).

\subsection{Magnetic Properties}

By construction of the present MF decoupling there is no sub-lattice
magnetization. Moreover since there is no anti-ferromagnetic term
in the Lagrangian the usual d-wave solutions of the $t-J$ model are
absent. Indeed, for two dimensions, the solutions with smallest free
energy were found to be invariant under lattice rotation. However,
a collection of other solutions that break rotational symmetry of
the lattice where found which are not the minima of the free energy,
as shown in the Appendix. The Spin-Spin correlation $\av{\mathbf{S}_{r}.\mathbf{S}_{r'}}$
functions, at the MF level, are exponentially decaying with the distance
and negative for $r\neq r'$ at finite $T$ and for all non-trivial
phases. This is compatible with the RVB picture where there is singlet
formation for $r$ and $r'$ not nearest neighbors.

\subsection{Charge Properties}

Using that \begin{equation}
c_{\mathbf{k},\sigma}=\frac{1}{\sqrt{V}}\sum_{\mathbf{q}}\left(e_{\mathbf{q}}^{\dagger}s_{\mathbf{q}+\mathbf{k},\sigma}+\sigma s_{\mathbf{q},-\sigma}^{\dagger}d_{\mathbf{q}+\mathbf{k}}\right)\end{equation}
 we can write the Green's function as \begin{eqnarray}
 &  & \langle T_{\tau}c_{k,\sigma}(\tau)c_{k,\sigma}^{\dagger}\rangle=\frac{1}{V}\sum_{\mathbf{q}}\nonumber \\
 &  & \left\{ \av{\mathit{e}_{q}^{\dagger}(\tau)\mathit{e}_{q}}_{0}\av{\mathit{s}_{k+q,\sigma}(\tau)\mathit{s}_{k+q,\sigma}^{\dagger}}_{0}\right.\nonumber \\
 &  & +\av{\mathit{d}_{q}(\tau)\mathit{d}_{q}^{\dagger}}_{0}\av{\mathit{s}_{q-k,-\sigma}^{\dagger}(\tau)\mathit{s}_{q-k,-\sigma}}_{0}\nonumber \\
 &  & +\sigma\av{\mathit{d}_{q}(\tau)\mathit{e}_{-q}}_{0}\av{\mathit{s}_{q-k,-\sigma}^{\dagger}(\tau)\mathit{s}_{k-q,\sigma}^{\dagger}}_{0}\nonumber \\
 &  & \left.+\sigma\av{\mathit{e}_{-q}^{\dagger}(\tau)\mathit{d}_{q}^{\dagger}}_{0}\av{\mathit{s}_{k-q,\sigma}(\tau)\mathit{s}_{q-k,-\sigma}}_{0}\right\} \end{eqnarray}
 For the case with no pairing correlations between fermions and considering
the bosons to be condensed leads to a Fermi liquid-like behavior for
the electron Green's function where the pole like structure of the
coherent part is determined by the spectrum of the fermions and the
condensates give the Z renormalization factor. Since condensation
was not allowed our phase diagram apparently misses two phases of
Ref.\cite{Lee_2006} away from $T=0$: the normal Fermi liquid phase
and the superconducting (SC) phase. In one dimension the phase ($\chi\neq0,\ \Delta=0$)
has a charge (bosonic) gap that goes to zero with the temperature
and a vanishing fermionic gap (see Fig. \ref{fig:1D_FreeEn_T}). However,
in two dimensions both gaps are zero in this phase (see Fig. \ref{fig:2D_FreeEn_T}).
Therefore at least at zero temperature there will be Bose condensation
and assuming a small interlayer coupling we may have Bose condensation
at finite temperature. Here we consider that the condensation temperature
is smaller than the considered range of temperatures. Note that a
conductor state is found for the $\chi\neq0,\ \Delta=0$ phase as
seen ahead in the spectral function where the spectral weight goes
down to zero energy. For the other nontrivial phases a spin gap is
present at low temperatures and one expects an insulator state. This
is also corroborated by the spectral function analysis where no substantial
spectral weight is seen near zero energy.

No condensation implies that the {}``superfluid'' density $\rho_{a}(k)=V^{-1}\left|\av{a_{k}a_{k}}_{0}\right|$
($a=d,e$) is zero for all $k$ and thus no SC correlations are present.
Note that the symmetry of the SC gap is given by the SC correlation
functions of the $s$ particles. The superconductor correlation function,
at the MF level, is given by \begin{eqnarray}
 &  & \av{c_{k,1}^{\dagger}c_{-k+p,-1}^{\dagger}}=\sum_{K}\delta_{p,2K}\nonumber \\
 &  & \left[\av{s_{k-K,1}^{\dagger}s_{-k+K,-1}^{\dagger}}_{0}\frac{\av{e_{-K}e_{-K}}_{0}}{V}\right.\nonumber \\
 & - & \left.\av{s_{-k+K,-1}s_{k-K,1}}_{0}\frac{\av{d_{K}^{\dagger}d_{K}^{\dagger}}_{0}}{V}\right]\nonumber \\
 & = & 0\end{eqnarray}
 where $\delta_{k,0}$ is $2\pi$ periodic in $k$ so the sum over
$K$ is reduced to two factors. However for the phases where $\Delta\neq0$
the anomalous spin correlation functions (defined in Eq. \ref{eq:non_phys_corr})
are finite, which can be seen as a precursor for superconductivity.%

\section{Spectral function}

\begin{figure*}
\begin{tabular}{ll}
\begin{tabular}{l}
(a)\tabularnewline
\includegraphics[width=6cm]{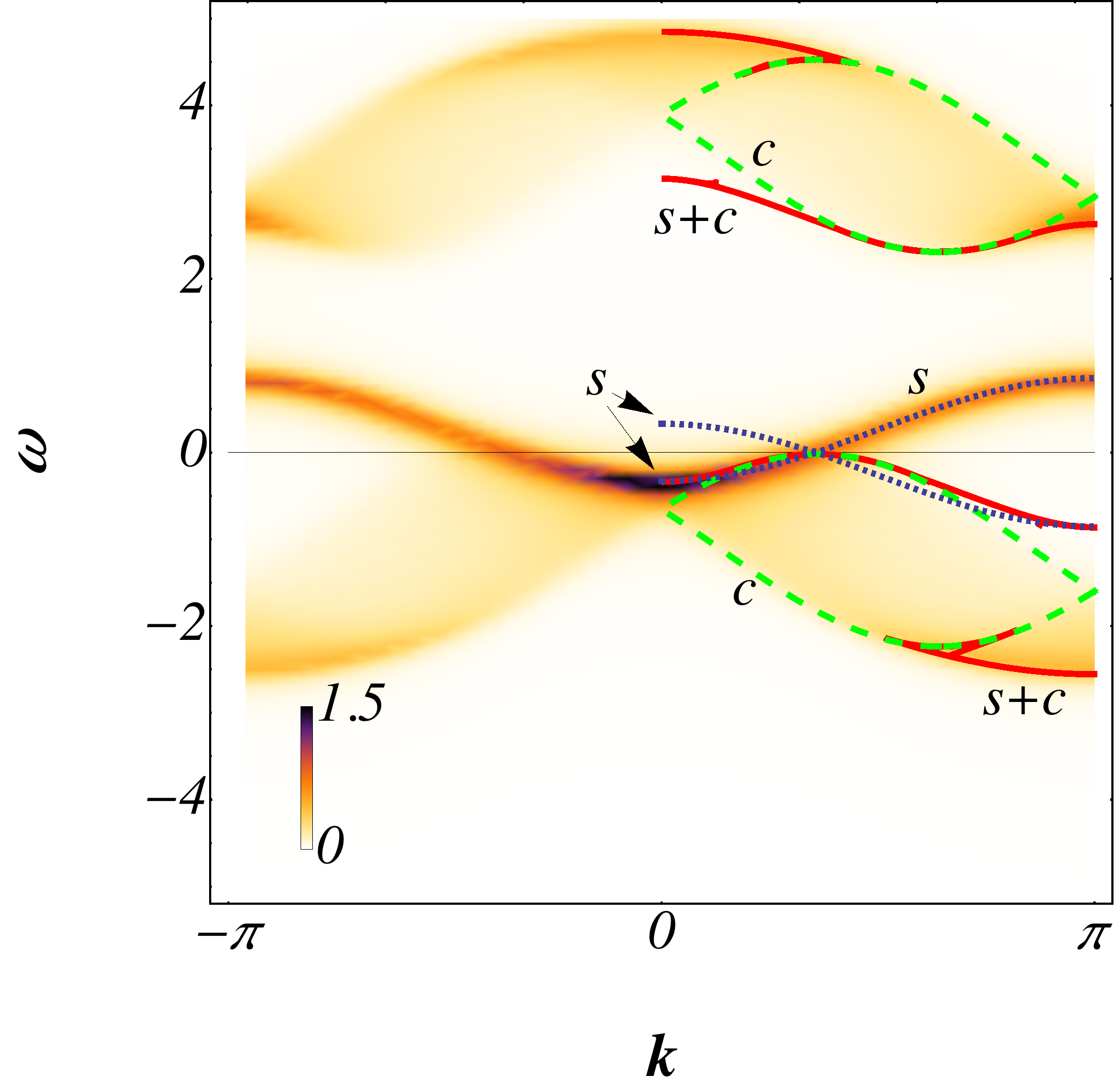}\tabularnewline
\end{tabular} & \begin{tabular}{c}
\includegraphics[width=3cm]{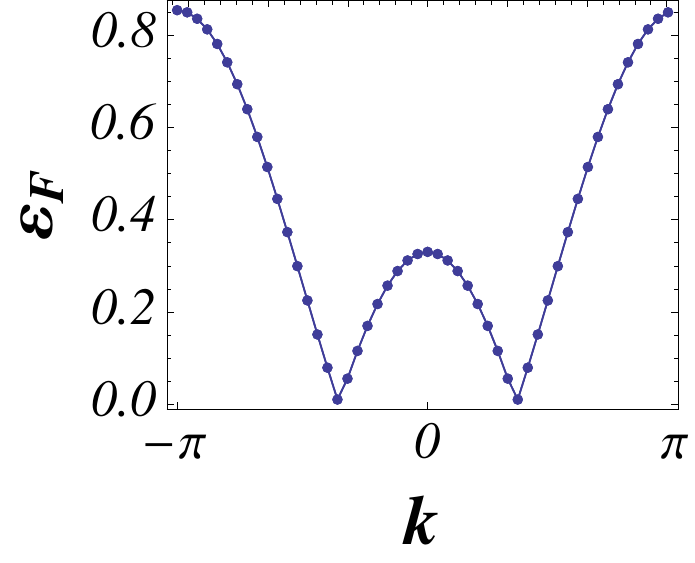}\tabularnewline
\includegraphics[width=3cm]{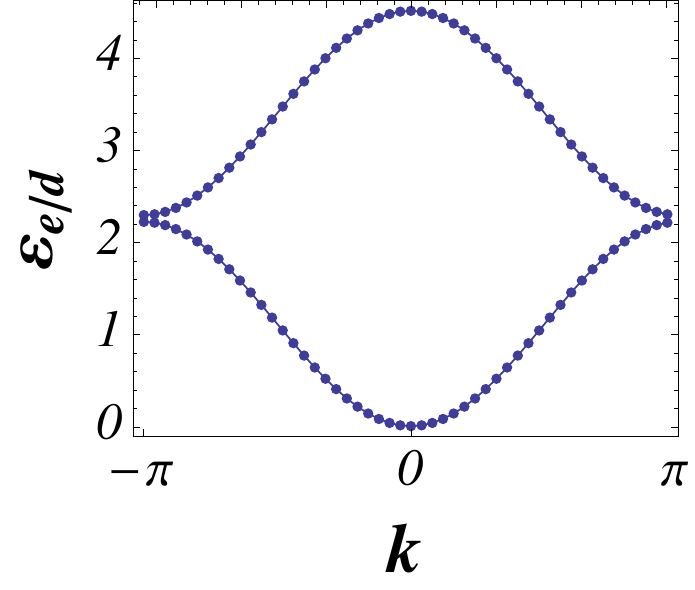}\tabularnewline
\end{tabular}\tabularnewline
\begin{tabular}{l}
(b)\tabularnewline
\includegraphics[width=6cm]{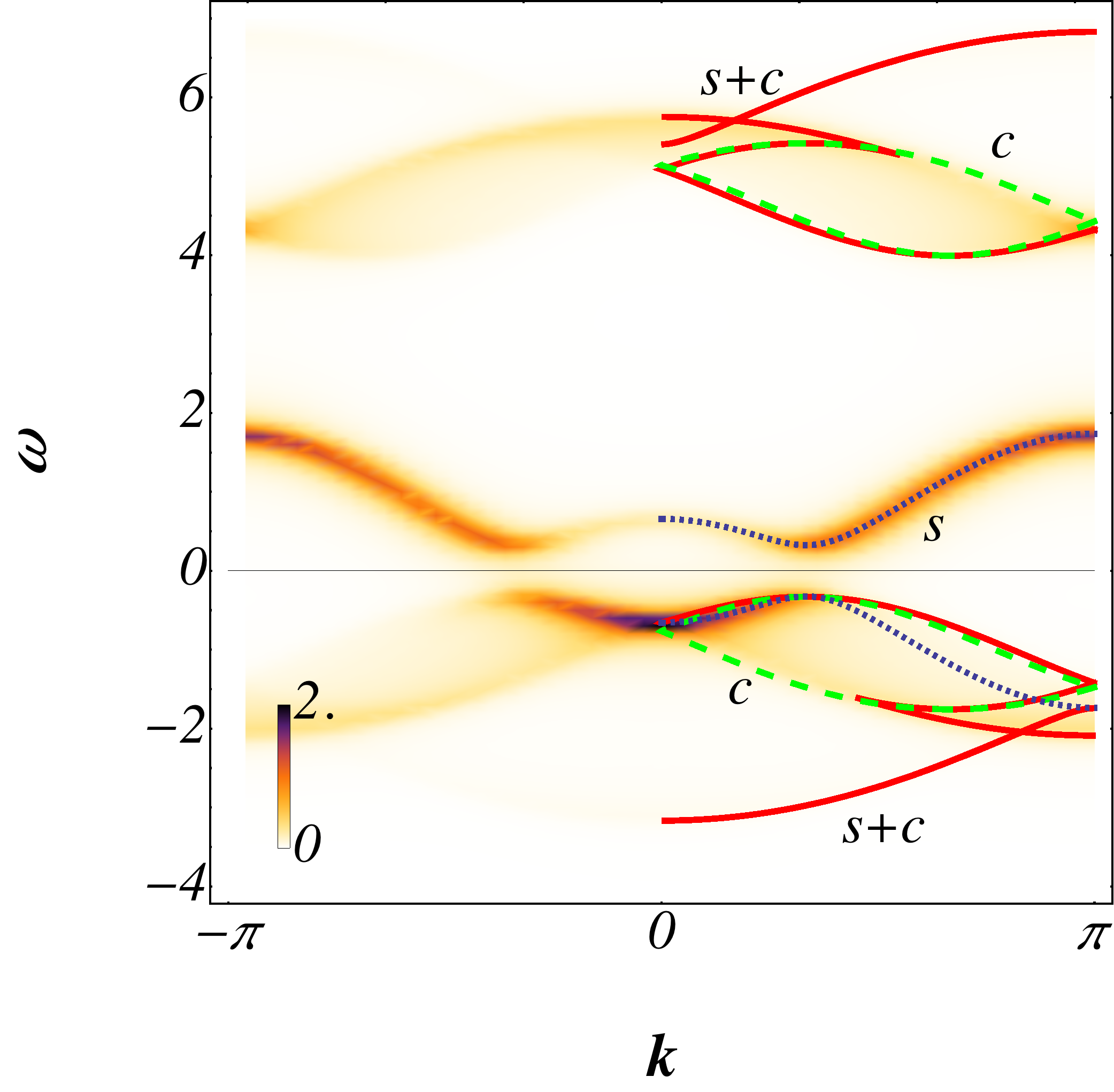}\tabularnewline
\end{tabular} & \begin{tabular}{c}
\includegraphics[width=3cm]{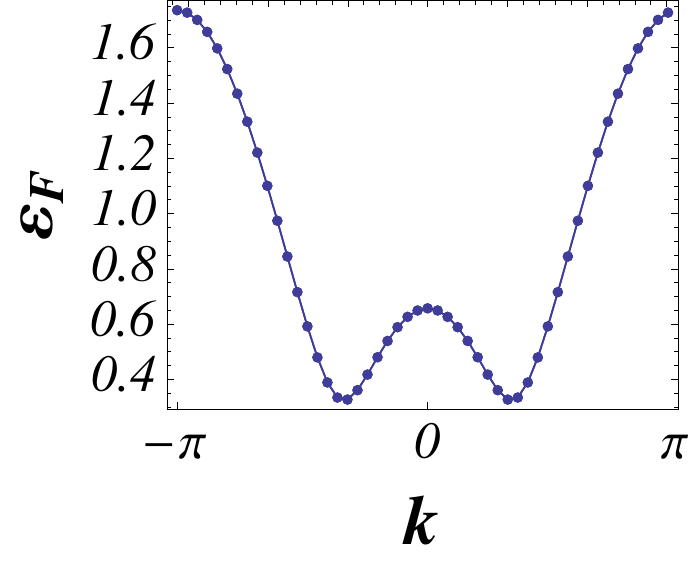}\tabularnewline
\includegraphics[width=3cm]{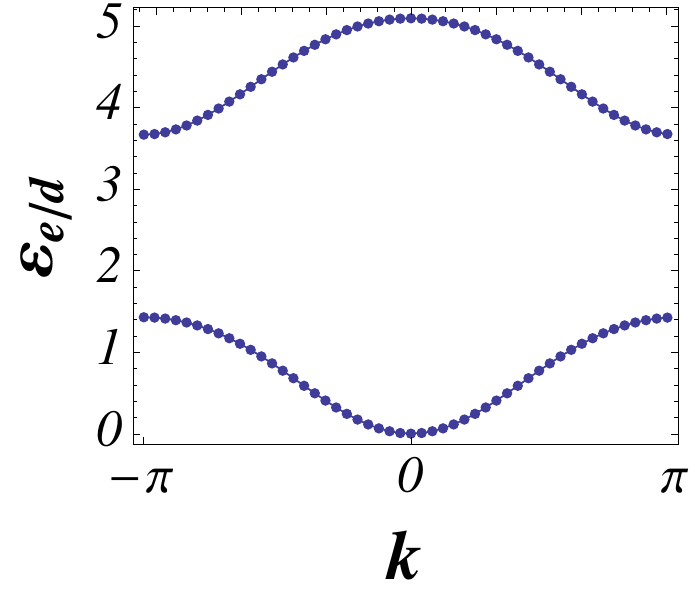}\tabularnewline
\end{tabular}\tabularnewline
\begin{tabular}{l}
(c)\tabularnewline
\includegraphics[width=6cm]{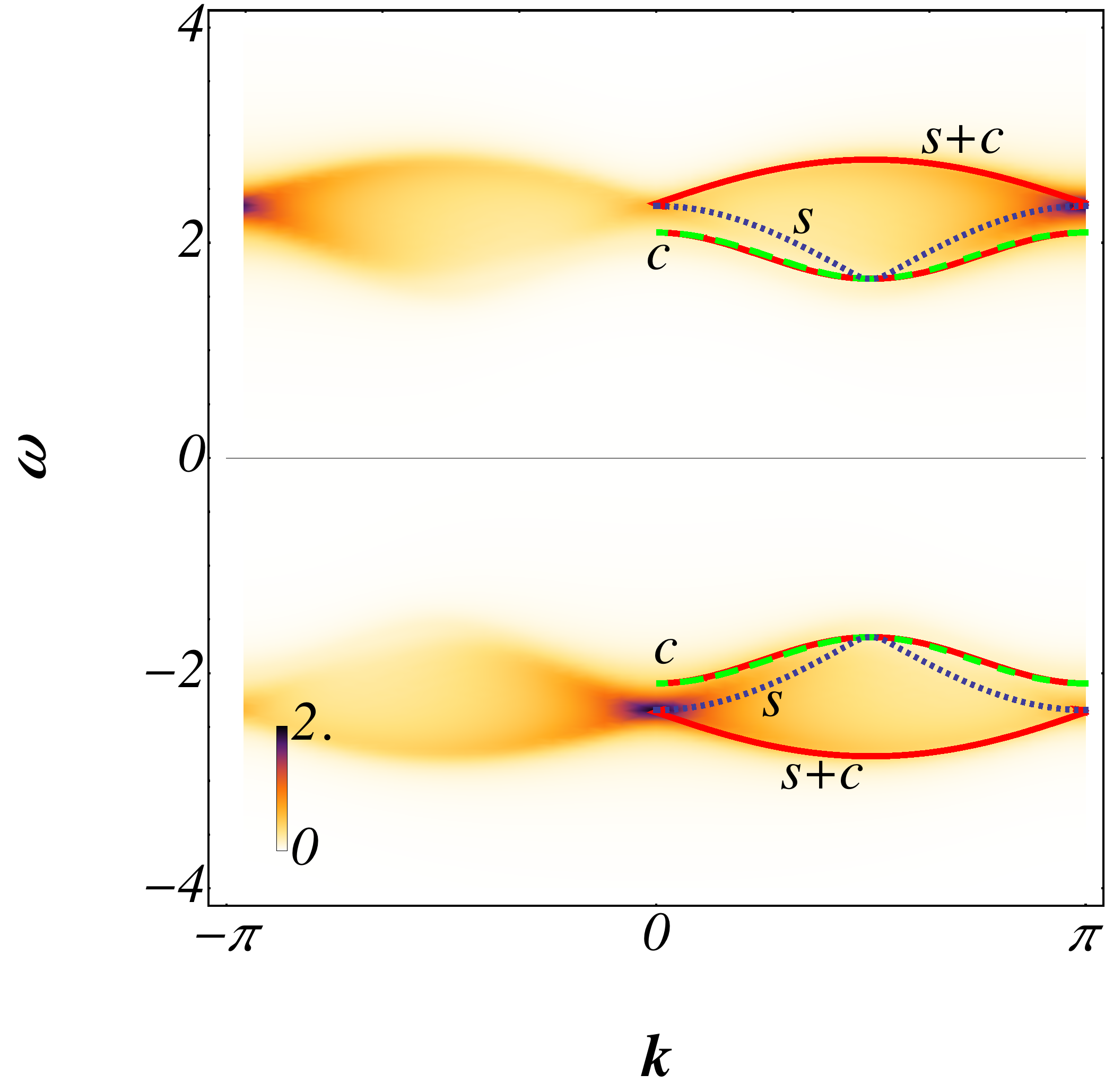}\tabularnewline
\end{tabular} & \begin{tabular}{c}
\includegraphics[width=3cm]{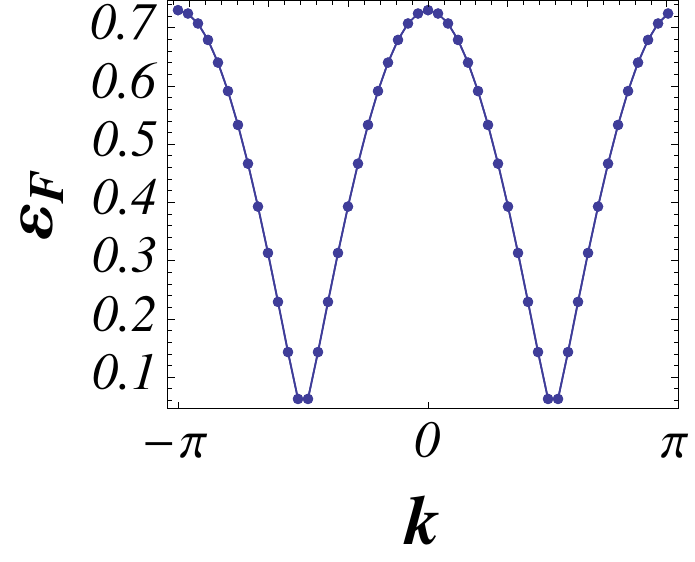}\tabularnewline
\includegraphics[width=3cm]{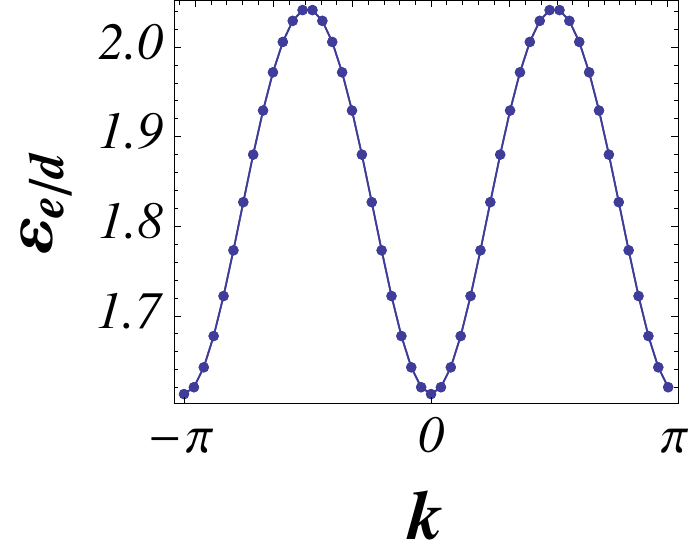}\tabularnewline
\end{tabular}\tabularnewline
\end{tabular}\caption{\label{fig:1D_Spectral_Function}Spectral function and single particle
energies for the Hubbatd chain:
(a) Phase $\chi\neq0;\ \Delta=0$ computed
for $U=4;\ x=0.3;\ T=0.06$. (b) Phase $\chi\neq0;\ \Delta\neq0$
computed for $U=4;\ x=0.5;\ T=0.06$. (c) Phase $\chi=0;\ \Delta\neq0$
computed for $U=4;\ x=0;\ T=0.06$. 
The $U-x-T$ parameters were chosen such that each solution is a minimum of the
free energy. Colored lines are plotted along regions of considerable
spectral weight and are obtained as follows: (Dark Blue, dotted line)
varying the momentum of the fermions while keeping the momentum of
the bosons to its minimal energy; (Green, dashed line) fixing the
fermion energy to its minimal value and changing the momentum of the
bosons; (Red, solid line) requiring that both excitations have equal
velocities. 
}

\end{figure*}

An efficient way to obtain information about the excitation spectrum
of a strongly correlated system is through the spectral function.
This is defined as the imaginary part of the Green's function and
is directly measurable through photoemission experiments. For the
theory considered here the electron spectral function at the MF level
can be written as\begin{widetext} \begin{eqnarray}
A_{\sigma}(\omega,k) & = & \frac{1}{V}\sum_{q}\left[\delta(\omega+\omega_{e,q}+\omega_{F,-q-k})\left[1+\mathit{n}_{b,e}(q)-\mathit{n}_{f}(-q-k)\right]\left|u_{F,q+k}v_{B,-q}+u_{B,-q}v_{F,q+k}\right|^{2}\right.\nonumber \\
 &  & +\delta\left(\omega-\omega_{d,q}+\omega_{F,q-k}\right)\left[\mathit{n}_{f}(q-k)+\mathit{n}_{b,d}(q)\right]\left|u_{B,q}u_{F,k-q}+v_{F,k-q}\bar{v}_{B,q}\right|^{2}\nonumber \\
 &  & +\delta\left(\omega-\omega_{d,q}-\omega_{F,k-q}\right)\left[1+\mathit{n}_{b,d}(q)-\mathit{n}_{f}(k-q)\right]\left|u_{F,k-q}v_{B,q}-u_{B,q}v_{F,k-q}\right|^{2}\nonumber \\
 &  & \left.+\delta\left(\omega+\omega_{e,q}-\omega_{F,q+k}\right)\left[\mathit{n}_{f}(q+k)+\mathit{n}_{b,e}(q)\right]\left|u_{B,-q}u_{F,q+k}-v_{F,q+k}\bar{v}_{B,-q}\right|^{2}\right]\end{eqnarray}
 \end{widetext} and has information about the excitation spectra
and their spectral weights.

\begin{figure*}[!t]
 \begin{tabular}{ll}
\begin{tabular}{c}
\includegraphics[width=12cm]{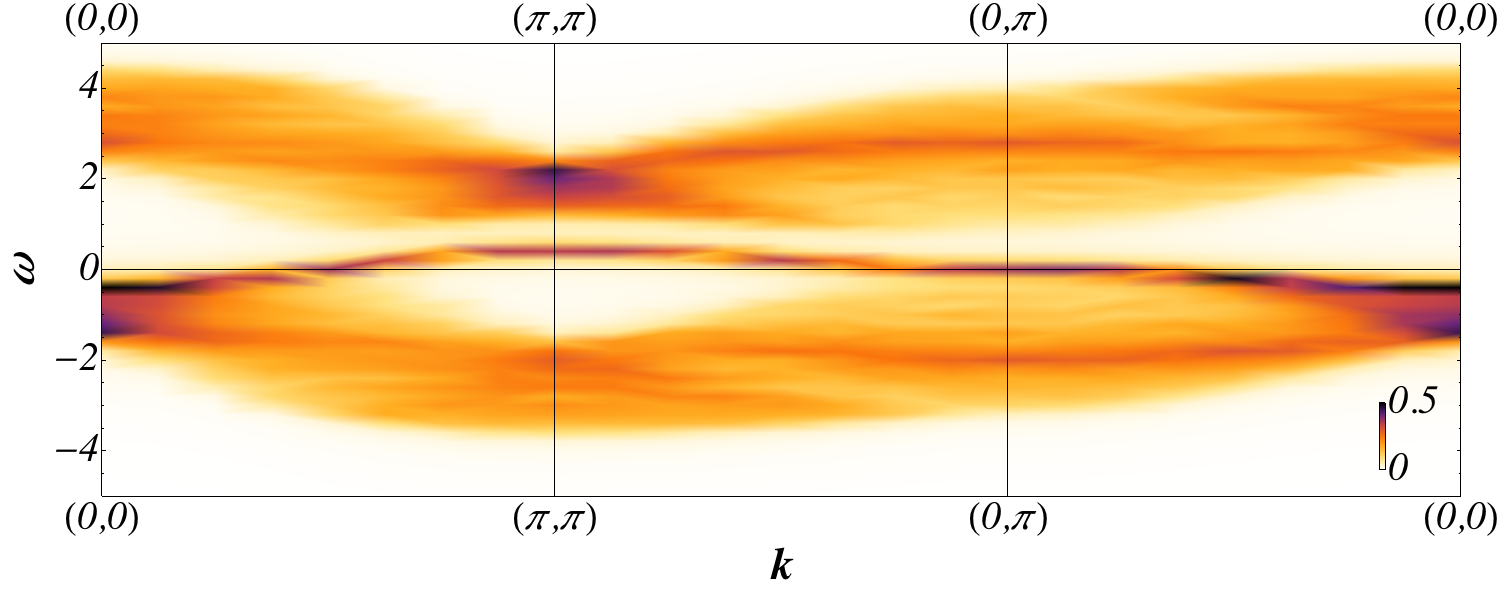}\tabularnewline
\end{tabular} & \begin{tabular}{c}
\includegraphics[width=4cm]{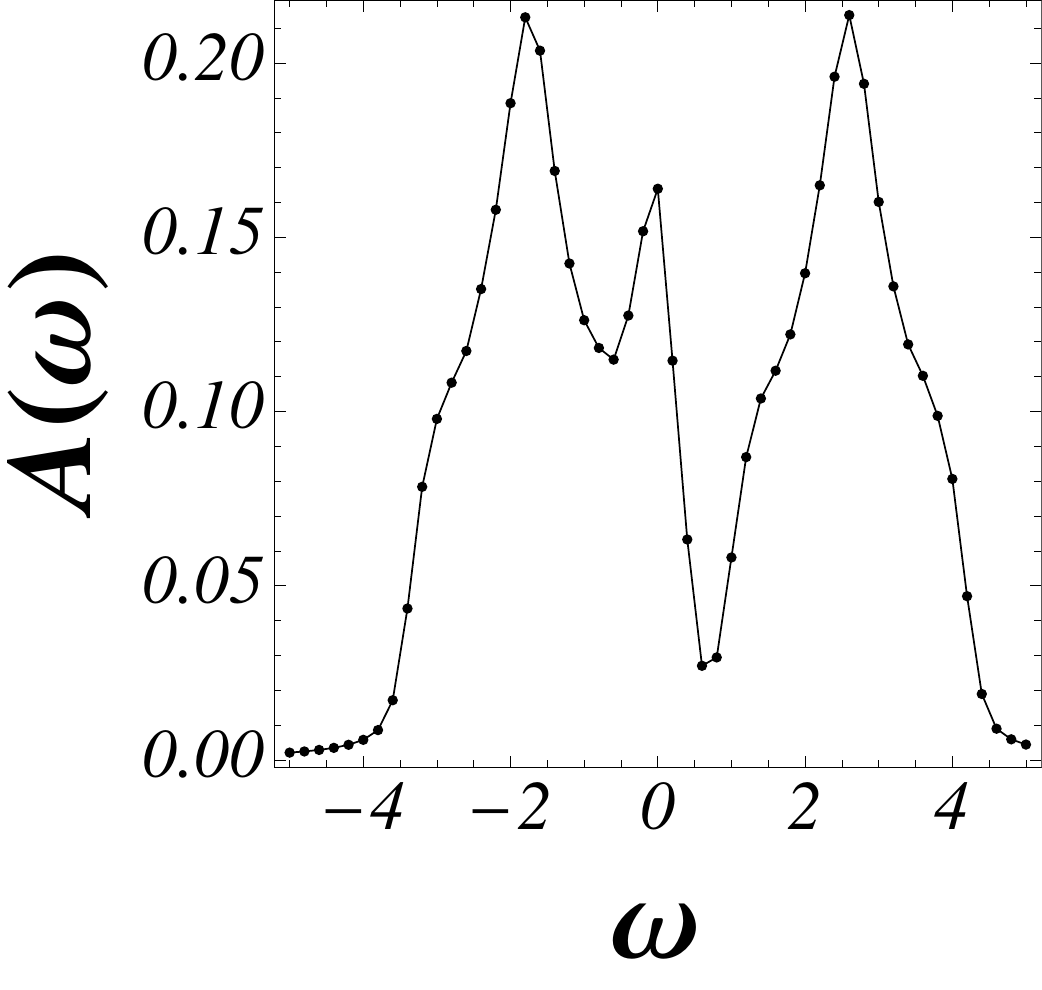}\tabularnewline
\end{tabular}\tabularnewline
\begin{tabular}{c}
\includegraphics[width=12cm]{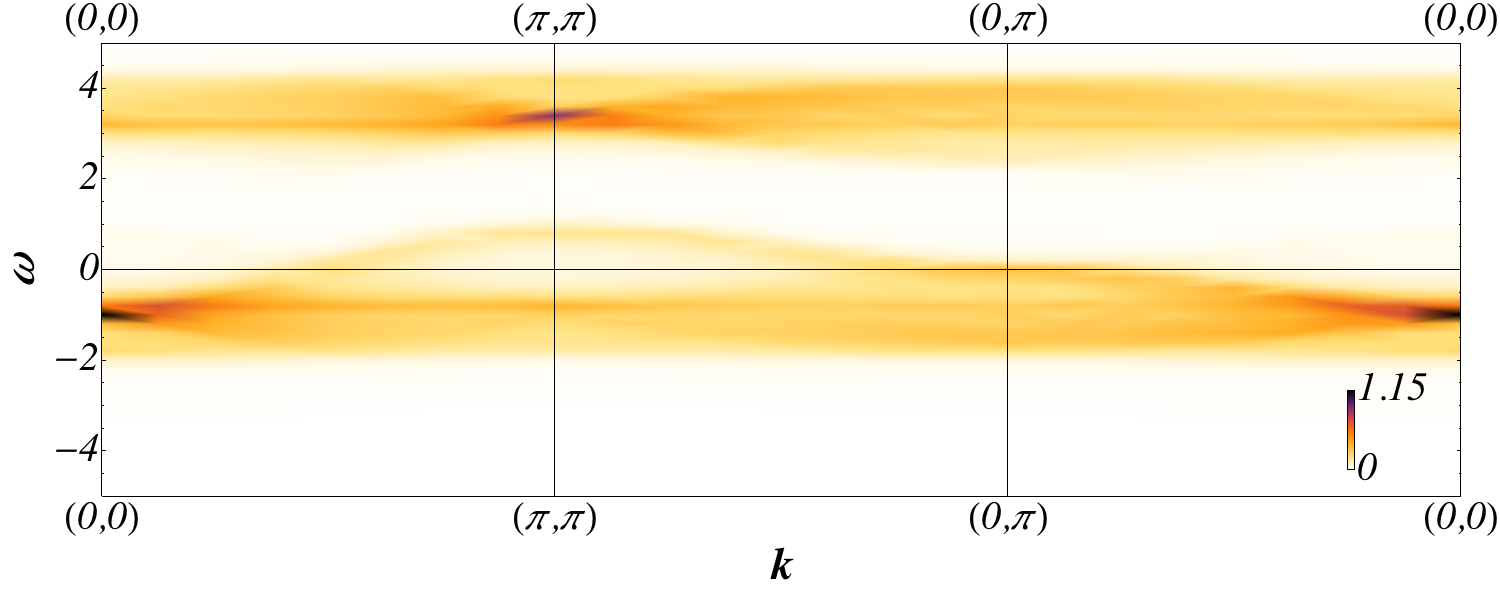}\tabularnewline
\end{tabular} & \begin{tabular}{c}
\includegraphics[width=4cm]{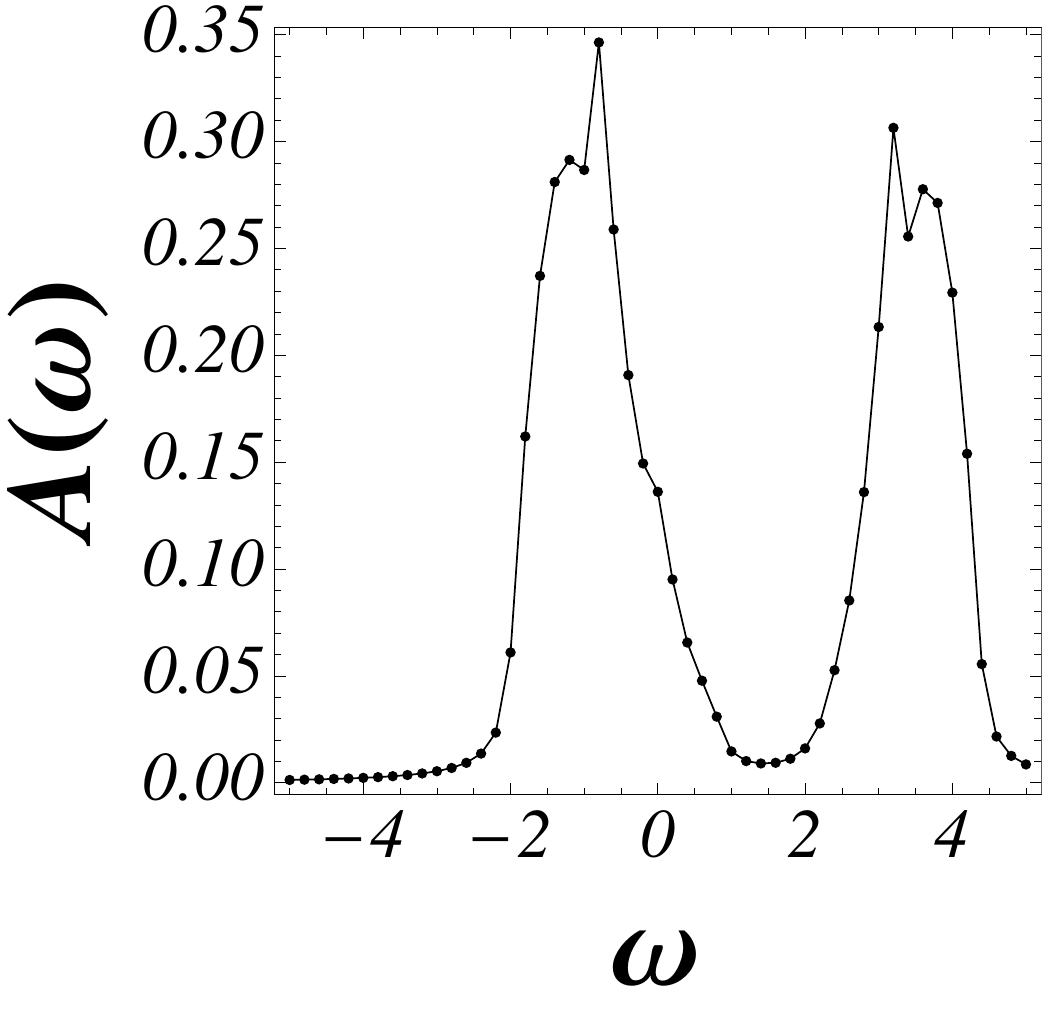}\tabularnewline
\end{tabular}\tabularnewline
\end{tabular}\caption{\label{fig:2D_Spectral_Function}Spectral function and density of
states computed at the MF level for a square lattice $16\times16$.
Upper panel: phase $\chi\neq0,\Delta=0$ ($U=4,T=0.01,x=0.1$). Lower
panel: phase $\chi\neq0,\Delta\neq0$ ($U=4,T=0.2,x=0.1$)}

\end{figure*}

\begin{figure}[!th]

\begin{centering}
\begin{tabular}{l}
\includegraphics[width=7cm]{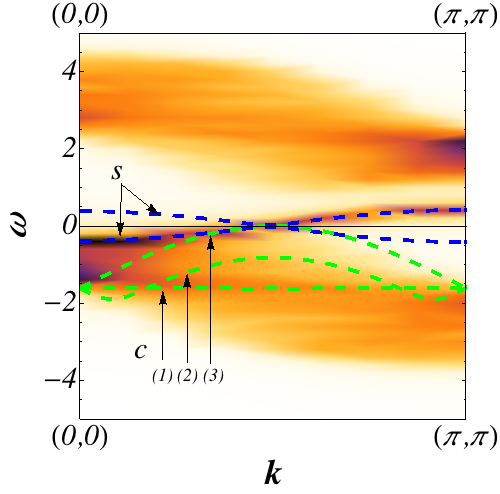}\tabularnewline
\end{tabular}
\par\end{centering}

\caption{\label{fig:2D_Spectral_Function-1}Spectral function for the square
lattice for the phase $\chi\neq0,\Delta=0$ along the nodal line.
The lines green are obtained fixing the fermion momentum at the Fermi
level along the directions $(1)=(\pi,0);\ (2)=(\pi/2,\pi/2);\ (3)=(0,\pi)$
and varying the momentum of the bosons. The blue lines are obtained
fixing the bosons energy to be minimal (corresponding to zero momentum)
and varying the fermionic momentum. }

\end{figure}

Experiments for one-dimensional insulators and conductors have shown
the fractionalization of the electronic degrees of freedom inside
the strongly correlated system. For instance experimental results
for the one-dimensional conductor TTF-TCNQ have been interpreted using
the Bethe ansatz solution showing clearly the traits of the spinon
and holon branches associated with the spin and charge degrees of
freedom (see Fig. 9 in \cite{Sing_2003} and Figs. 1 in \cite{Carmelo_2004_b,Bozi_2008}).
In the context of the high-temperature superconductors similar results
have been obtained for the spectral function in \cite{Graf_2007,Damascelli_2003}.
These results have also been interpreted in terms of some fractionalization
of the degrees of freedom in a way similar to the one-dimensional
case.

Fig. \ref{fig:1D_Spectral_Function} shows the spectral function for
the one-dimensional case computed at the MF level for the three non-trivial
MF solutions. The regions of larger spectral weight correspond to
three different processes. Two of those processes can be interpreted
as the fractionalization of the electron since the fermions or the
bosons (Blue or Green respectively) are fixed to the minimum value
of their energy and the other species is allowed to move along its
energy band. The high spectral weight of such regions is also observed
in the spectral function computed based on the exact solution of the
Hubbard model in 1D (REF). These are the so-called spectral lines
\textquotedbl{}c\textquotedbl{} and \textquotedbl{}s\textquotedbl{}
in refs. \cite{Sing_2003,Carmelo_2004_b,Bozi_2008}. The third process
can be interpreted as propagation of an \textquotedbl{}electron-like\textquotedbl{}
degree of freedom (Red) since it corresponds to a boson and a fermion
with the same velocity propagating together. These processes typically
define the boundary lines of the region with a (nearly) non-vanishing
spectral weight.

Consider first the top panel of Fig. \ref{fig:1D_Spectral_Function}
and the negative energy region corresponding to photoemission. In
the regions of higher spectral weight at low energies we see that
the lowest energy branch (s branch) is obtained fixing the bosons
at their lowest energy and changing the fermionic (spin) particles
along their bands. It is therefore a spin branch. Below this line
at higher energy (recall that by definition $\omega<0$ for photoemission
and therefore higher energies are more negative) there is a line that
is obtained fixing the fermion at the Fermi surface and changing the
boson energy along its band. It is therefore a charge (holon) band
(c branch). These two lines are clearly separated as in the exact
description from the Bethe ansatz and the experimental results \cite{Bozi_2008}.
They merge at the Fermi level as in the exact solution. Note that
in the region of high spectral weight there is a contribution from
a line obtained taking the velocities of the fermions and bosons as
equal. This means an \textquotedbl{}electronic-like\textquotedbl{}
excitation. For larger momenta spin and a charge branch also emerge
as in the exact solution. We also show the inverse photoemission spectra
(positive energies) including the lowest and the upper Hubbard bands.
In this phase there is a finite spectral weight at the Fermi energy
which implies a conducting phase. This is consistent with the gapless
fermionic band and the nearly gapless bosonic lowest band.

In the middle panel we consider the phase where both the hoppings
and gap functions are finite (White phase). In this case the spins
have a gap and the bosons are nearly gapless. The spectral function
has now a pseudogap at the Fermi level since there is a small spectral
weight. The finite energy structure is however quite similar to the
fully conducting phase. Note that another contribution to this region
is obtained fixing the fermion energy to its minimal (finite) value
and changing the momentum of the bosons (nearly gapless). Note again
that in the region of high spectral weight there is again a contribution
from a line obtained taking the velocities of the fermions and bosons
as equal.

\begin{figure*}
\begin{centering}
\begin{tabular}{l}
\includegraphics[width=0.8\textwidth]{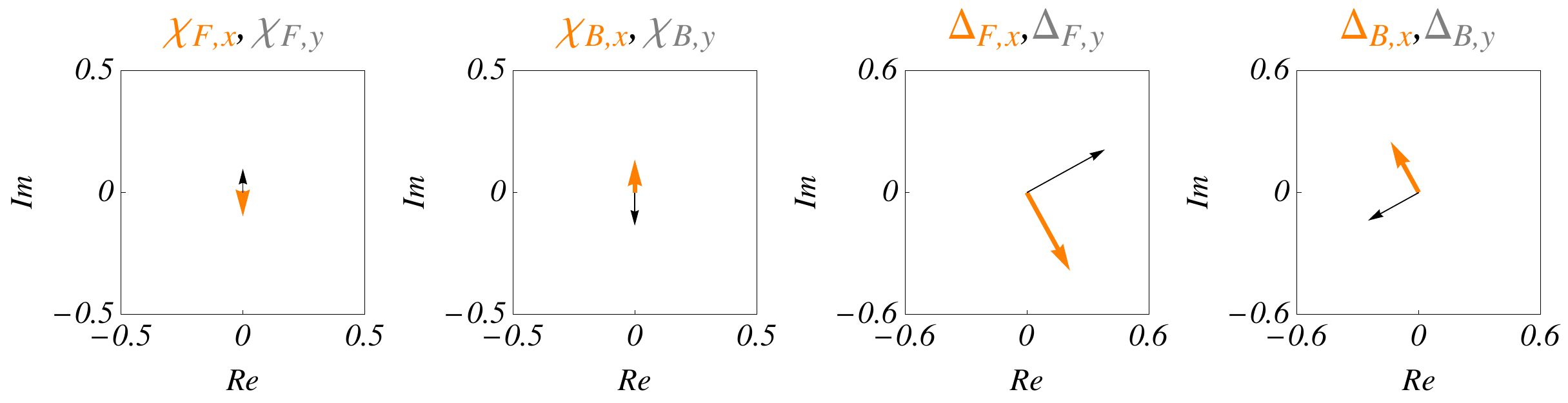}\tabularnewline
\end{tabular}
\par\end{centering}

\caption{\label{fig:example} This figure shows how the results displayed in
Fig. \ref{fig.MF_Sol} should be interpreted. }

\end{figure*}

Finally, in the lower panel we consider the Orange solution where
the hopping parameters vanish and the gap functions are finite. In
this case both the fermions and the bosons are gapped and the spectral
function has a large gap. Also, the results are presented at half-filling
and therefore the system is an insulator, as expected from the exact
solution.

The spectral function and density of states \begin{equation}
A(\omega)=\frac{1}{2}\frac{1}{V}\sum_{k,\sigma}A_{\sigma}(\omega,k)\end{equation}
 for the square lattice are shown in Figs.\ref{fig:2D_Spectral_Function}
and \ref{fig:2D_Spectral_Function-1}. For the $\chi\neq0,\Delta=0$
(Red) phase it presents the same qualitative features as the one dimensional
case. Namely a strong spectral weight is observed due to excitations
corresponding to an empty site at $k=0$ and a spinon that carries
the momentum of the physical electron (s branch). Excitations for
which the spinon is taken at the Fermi surface and the holon carries
the difference of momenta presents also some spectral weight (Green
lines of Fig.\ref{fig:2D_Spectral_Function-1}). Fig.\ref{fig:2D_Spectral_Function-1}
shows Green lines (c branches) corresponding to the spinon momentum
at the Fermi surface in the direction $(0,\pi),(\pi,\pi),(\pi,0)$
directions. The line reaching the Fermi energy corresponds to a fermion
with momentum in the $(0,0)-(\pi,\pi)$ segment. This scenario of
two spectral lines leaving the Fermi-surface of the spinons is also
obtained by other recent approaches \cite{Phillips_2009}. The results
for the White phase show a pseudogap structure as shown in the density
of states $A(\omega)$.

\section{Discussion}

We have extensively explored translationally invariant MF solutions
of the Hubbard model for a one dimensional chain and for a square
lattice using an electron representation introduced in \cite{Zou_1988}
and a MF decoupling in terms of link variables. In two dimensions
this includes non-trivial symmetries of the MF solutions as well as
nematic (translationally invariant but not rotationally invariant)
and quasi-1D phases (but no flux phases). Despite all the freedom
in the choice of solutions, the ones that minimize the free energy
were found to be invariant under lattice rotations. However, in some
phase space regions, the free energy difference between these less
conventional phases (non rotationally invariant) and the symmetric
ones (rotationally invariant) where found to be quite small signaling
a possible stabilization of such phases by some extra coupling in
the Hamiltonian. The symmetric phases where classified according to
the MF order parameters and their physical properties where obtained.

Generically we found a gapped phase for both fermionic and bosonic
(spinons and holons) degrees of freedom for zero doping. Contrarily
to the 2D case, where this phase was only found for half filling,
in the 1D case this phase extends to finite doping at finite temperatures
and its size in the $T-x$ phase diagram decreases with $U$. This
is in no contradiction with the exact solution for the ground state
since for zero temperature only at half filling have we found this
phase; the system is a conductor away from half filling. Although
no magnetic order is obtained from the present MF ansatz this state
is clearly a Mott insulator as one expects at half filling.

A conducting phase was found to be dominant for small and moderate
doping from zero to quite high temperatures presenting the qualitative
features of a RVB state. For the one dimensional conducting phase
one can identify some of the features of the spectral function of
Fig. \ref{fig:1D_Spectral_Function} with the ones obtained from the
exact Bethe Ansatz solution. In particular the lines carrying the
most part of the spectral weight can be identified with three kinds
of excitations. Two of them corresponding to a creation or annihilation
of slave particles with different velocities. It is tempting to interpret
such lines with fractionalization of the initial degrees of freedom.
In the two dimensional case these lines are also obtained although
they are not so clearly defined, in the sense that the spectral weight
is not very pronounced for the c branches. Near the Fermi-energy such
two lines are compatible with the description in \cite{Phillips_2009}.
Lines where both slave particles have equal velocities are possible
to draw only in the one dimensional case since they correspond to
a surface in 2D. In this case they typically represent boundaries
for the spectral weight. As in the former case one can try to interpret
these lines as \textquotedbl{}electron-like\textquotedbl{} particles
since they represent states where a slave boson and slave fermion
travel together with the same velocity. In the region where the $\chi\neq0,\Delta\neq0$
phase appears the temperature is larger than the spin gap, the charge
gap being zero. 
We note that in the one-dimensional case in the conducting phases the fermions are gapless
and the bosons are gapped while in the square lattice
the fermions are gapped and the bosons are gapless. 

The results presented here describe the finite-energy finite-temperature
phase diagram. At very low temperature a low energy theory where the
lowest energy bosonic branch is condensed and the higher energy bosonic
branch is frozen, will be presented elsewhere. It is particularly
interesting in this low energy regime to consider frustrated lattices
where it is expected that fractionalization will appear for either
a conducting system, or for insulating systems where the possibility
of spin liquids has been proposed.
\begin{acknowledgments}
We acknowledge partial support from Project PTDC/FIS/70843/2006. PR
acknowledges support through FCT BPD grant SFRH/BPD/43400/2008. Also
we acknowledge discussions with V.R. Vieira, P.A. Lee and Z. Tesanovic
at the early stages of this work. 
\end{acknowledgments}
\appendix

\section{Mean Field Solutions\label{sec:MF_sol}}

\begin{figure*}
\begin{tabular}{ccc}
\includegraphics[width=0.95\columnwidth]{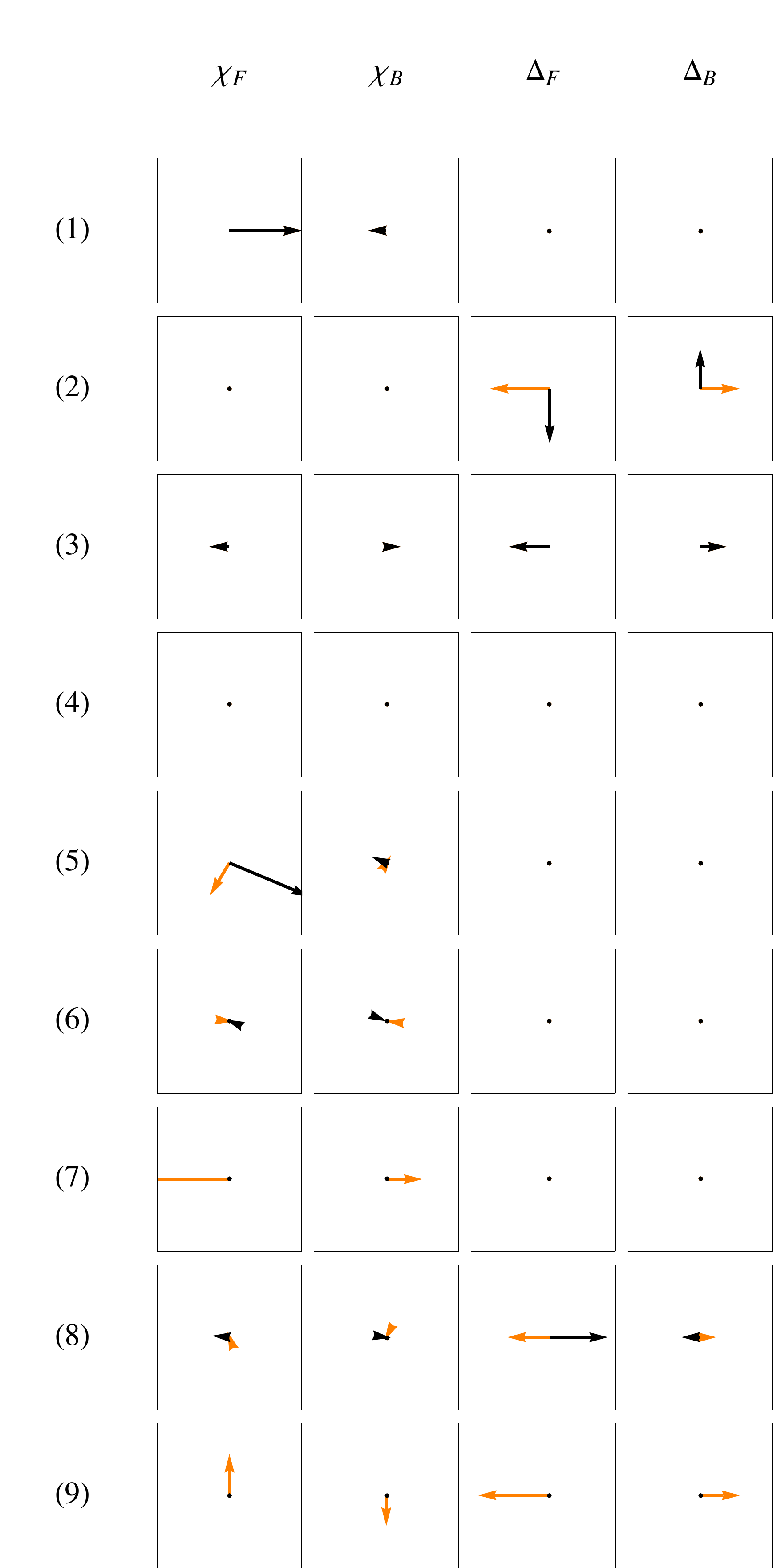}  & \includegraphics[width=0.95\columnwidth]{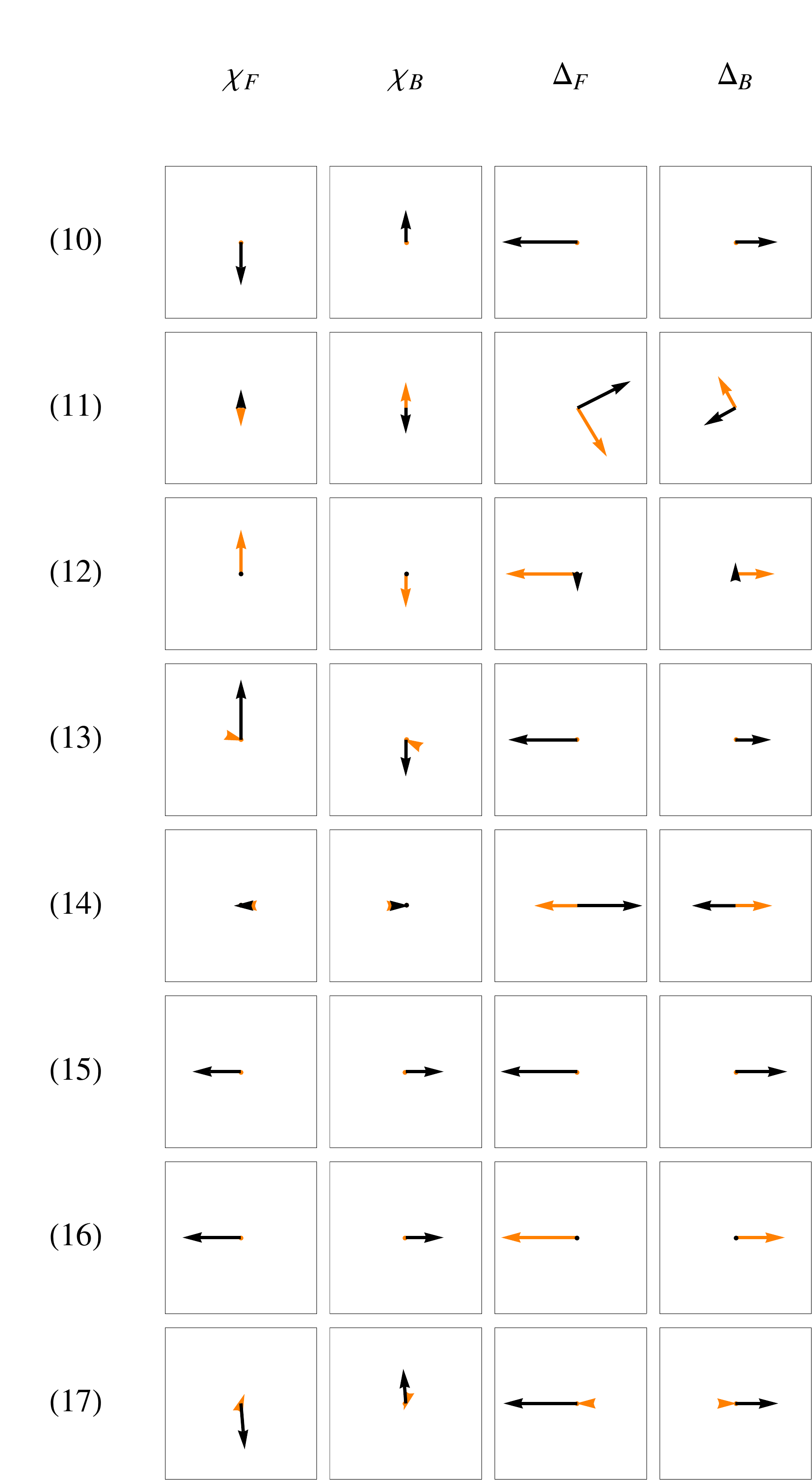}  & \tabularnewline
\end{tabular}

\caption{\label{fig.MF_Sol}Mean Field Solutions }

\end{figure*}

In this section we present a table with some of the solutions found
for the square lattice, solving the MF equations. These solutions
have low values of the free energy even though not absolute minima. 

The results presented in Fig. \ref{fig.MF_Sol} are obtained for $U=4,x=0.2,T=0.1$
for the two dimensional square lattice and are representative of the
types of solutions found in other points of phase space. Solutions
$(1),(3)$ and $(4)$ are the ones considered in the main text since
they alternatively minimize the free energy depending on the considered
phase space region: $(1)$ corresponds to the $\chi\neq0,\Delta=0$
phase, $(3)$ corresponds to the $\chi\neq0,\Delta\neq0$$ $ phase
and $(4)$ is the incoherent phase where $\chi=0,\Delta=0$. Note
that the solution $(2)$ referred in the main text corresponds to
the zero doping limit of solution $(3)$, the one shown in Fig. \ref{fig.MF_Sol}
is also a solution of the type $\chi=0,\Delta\neq0$ but it never
presents the lowest free energy. The table entries are complex plane
plots representing the MF parameters: columns 2 to 5 represent the
values of $\chi_{F/B,\delta},\Delta_{F/B,\delta}$ for $\delta=\hat{e}_{x}$
(orange arrow) and $\delta=\hat{e}_{y}$ (black arrow); column 6 displays
the values of $\mu$ (black arrow) and $i\lambda$ (orange arrow).
The scales of the axes are shown in Fig. \ref{fig:example}. We note
some remarkable solutions obtained here: solution $(4)$ is an anisotropic
$\chi\neq0,\Delta=0$ solution; solutions $(9)$ and $(10)$ are related
by a lattice rotation and represent one dimensional-like correlations
where in one of the directions the MF parameters are zero; solution
$(16)$ has $\chi\neq0,\Delta=0$ in the $x$ direction and $\chi=0,\Delta\neq0$
in the $y$ direction.

\bibliographystyle{apsrev}

\end{document}